\documentclass[twocolumn,amsmath,amssymb,superscriptaddress,nofootinbib,longbibliography,pra]{revtex4-2}
\usepackage{graphicx}% Include figure files
\usepackage{dcolumn}% Align table columns on decimal point
\usepackage{braket}

\renewcommand\vec{\mathbf}

\usepackage[version=3]{mhchem} % Formula subscripts using \ce{}

\begin{document}

\title{Polariton transport in 2D semiconductors: Phonon-mediated transitions between ballistic, superdiffusive and exciton-limited regimes} 

\author{Jamie M. Fitzgerald}
\email{jamie.fitzgerald@physik.uni-marburg.de}
\affiliation{Department of Physics, Philipps-Universität Marburg, 35032, Marburg, Germany}

\author{Roberto Rosati}
\affiliation{Department of Physics, Philipps-Universität Marburg, 35032, Marburg, Germany}

\author{Ermin Malic}
\affiliation{Department of Physics, Philipps-Universität Marburg, 35032, Marburg, Germany}

\date{\today}

\begin{abstract}
Exciton transport in 2D semiconductors holds promise for room-temperature, ultra-compact optoelectronic devices, but it is limited by short propagation distances. Hybridization of excitons with cavity photons to form exciton-polaritons can enhance the propagation by orders of magnitude, enabling a coherent, ballistic transport. However, a microscopic understanding of the role of phonons is still lacking, particularly regarding their influence on the crossover from the ballistic to the diffusive polariton transport regime. Here, we investigate the spatiotemporal polariton dynamics in \ce{MoSe2} monolayers at moderate to high temperatures, explicitly including the phonon-mediated coupling to the intervalley exciton reservoir. We identify three distinct transport regimes: (i) an initial sub-ps ballistic-like regime characterized by a phonon-induced velocity renormalization, (ii) a transient, few-ps superdiffusive regime characterized by strongly enhanced diffusion, and (iii) a slower, exciton-limited diffusion following thermalization. The gained microscopic insights will trigger and guide future experimental studies on the phonon-mediated polariton transport in atomically thin semiconductors.
\end{abstract}

\maketitle

\section*{Introduction}

Monolayers of transition metal dichalcogenides (TMDs) support tightly bound Wannier excitons that are stable at room temperature and interact strongly with light \cite{mak2010atomically,wang2018colloquium,mueller2018exciton}. Given their charge neutrality, two-dimensional nature, and quasi-bosonic statistics, there has been long-standing interest in the in-plane transport properties of TMD excitons for both fundamental many-body physics and technological applications \cite{perea2022exciton,malic2023exciton}. In particular, exciton transport plays a key role in light-energy conversion, including photovoltaics \cite{jariwala2017van}, light-emitting diodes \cite{ross2014electrically}, and photosynthesis \cite{bredas2017photovoltaic}. There is also great potential for fast and energy-efficient excitonic circuits that bridge nano-scale electronics and micron-scale photonics \cite{unuchek2018room}. 

At elevated temperatures and low excitation power, a spatially localized population of excitons spreads isotropically via conventional phonon-driven diffusion, as described by Fick's law \cite{perea2022exciton}. The resulting photoluminescence (PL) from exciton recombination can be used to track exciton density in both time and space \cite{mouri2014nonlinear, kato2016transport, kulig2018exciton, uddin2020neutral, wagner2021nonclassical}, while dielectric \cite{carmesin2019quantum,li2021dielectric} and strain engineering \cite{branny2017deterministic,darlington2020imaging} have been demonstrated to direct the flow of exciton current. The rich exciton landscape of TMDs \cite{malic2018dark} is fundamental to understanding exciton diffusion, since the relative position of bright and dark exciton states is sensitive to the dielectric environment \cite{zipfel2020exciton} and strain \cite{rosati2020strain,kumar2024strain}, as demonstrated by exciton anti-funnelling observed in tungsten-based TMDs \cite{rosati2021dark}.
Furthermore, vertical stacking of heterostructures gives rise to spatially separated and long-lived interlayer excitons with permanent out-of-plane dipole moments \cite{sun2022excitonic}. This allows for electric field \cite{unuchek2018room,tagarelli2023electrical} and twist-angle \cite{li2021interlayer} control of exciton diffusion. Despite these notable attributes, exciton transport is inherently slow, with typical diffusion coefficients on the order of $1$-$10$ cm$^2$/s in TMD monolayers \cite{kulig2018exciton}. 

\begin{figure}[t!]
\includegraphics[width=\columnwidth]{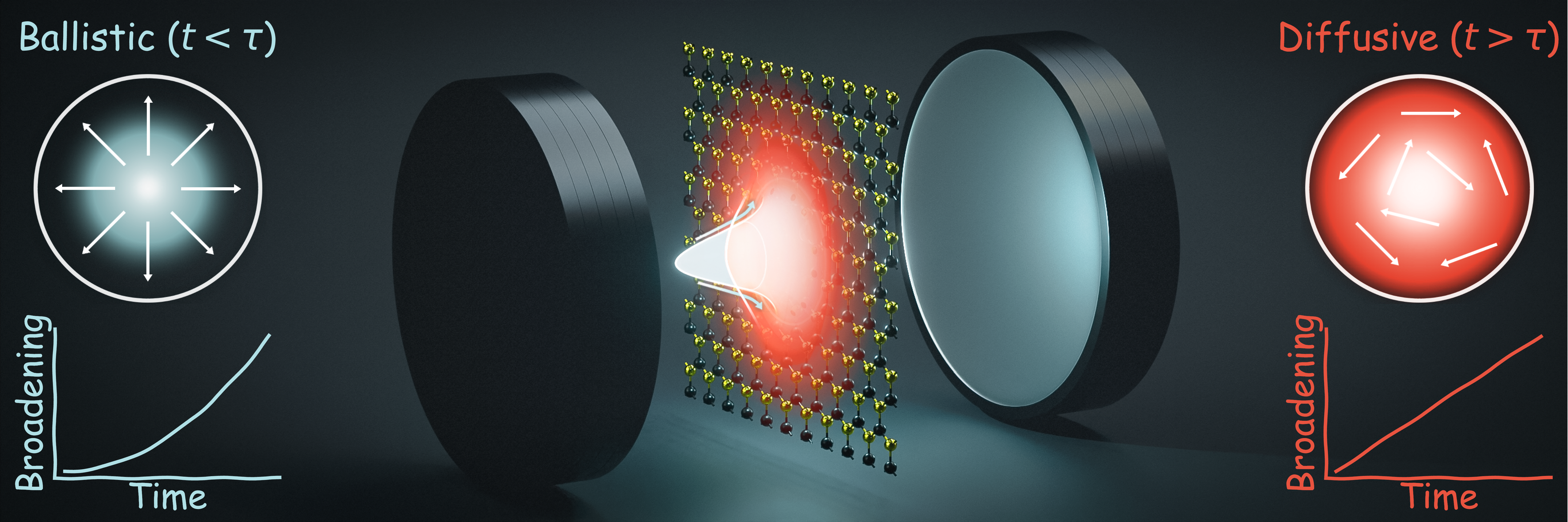}
\caption{(a)~Illustration of a \ce{MoSe2} monolayer integrated within a Fabry-P\'{e}rot microcavity. The system is in the strong-coupling regime, where transport behaviour is heavily modified due to the large group velocities of exciton-polaritons. The ballistic (left, blue) and diffusive (right, red) regimes are depicted, with the corresponding behaviour of the real-space broadening in time. Here, $\tau$ represents a typical thermalization time of the system and dictates the crossover from rapid ballistic-like expansion (quadratic) to a slower diffusive regime (linear).
\label{fig:fig_1}}
\end{figure}

In contrast, photons are ideal information and energy carriers that can propagate over macroscopic length scales in materials at a significant fraction of the speed of light. However, developing compact integrated photonic circuitry necessitates strong non-linearities at low optical intensities \cite{espinosa2013complete}. Exciton-polaritons are hybrid light-matter states that promise the best of both worlds, inheriting a light effective mass from their photon constituent, and tunability and nonlinearity from their material component \cite{schneider2018two}. They are also more resistant to disorder than pure excitons due to the motional narrowing effect \cite{wurdack2021motional}. Over the past decades, exciton polariton transport has been explored in conventional semiconductors in both microcavity \cite{ freixanet2000plane, steger2013long} and waveguide structures \cite{rosenberg2018strongly}. More recently, there has been intense focus on materials with a large Rabi splitting that can support long-range polariton propagation at room-temperature, such as organics \cite{lerario2017high,rozenman2018long,hou2020ultralong,balasubrahmaniyam2023enhanced,xu2023ultrafast}, perovskites \cite{su2018room,jin2023enhanced,black2024long,dang2024long}, and TMDs \cite{wurdack2021motional,shan2021spatial,guo2022boosting,liu2023long,xie20252d}. In particular, recent pioneering experiments have reported a detuning-dependent crossover between ballistic and diffusive regimes in exciton polariton transport \cite{balasubrahmaniyam2023enhanced,xu2023ultrafast}. In the context of TMDs, the ballistic propagation of polaritons across tens of microns has been demonstrated, along with trapping and manipulation using a spatially dependent modification of the cavity length \cite{wurdack2021motional}. However, the role that dark intervalley excitons play in polariton transport is still largely unknown. In our previous work, the full exciton landscape was shown to be essential for understanding polariton optics \cite{ferreira2022signatures,ferreira2024revealing} and relaxation dynamics \cite{fitzgerald2024circumventing}. For the latter, intervalley KK$^\prime$ excitons were shown to act as additional exciton reservoir to efficiently populate polariton states and bypass the relaxation bottleneck \cite{tassone1997bottleneck}.

In this work, we focus on a representative hBN-encapsulated \ce{MoSe2} monolayer integrated within a Fabry-P\'{e}rot microcavity (Fig.~\ref{fig:fig_1}), and calculate the full spatiotemporal dynamics of exciton-polaritons by solving the polariton Boltzmann transport equation. Crucially, we include the momentum-resolved, real-space dynamics of both polaritons and dark intra- and intervalley reservoir excitons. This provides a microscopic description of the phonon-driven spatiotemporal relaxation of the system, which is explored for different temperatures and detunings. We observe three distinct regimes in the time evolution of the polariton cloud: First, there is a rapid ballistic-like expansion driven by the ultrafast relaxation of hot excitons into polaritons. This is followed by a transient superdiffusive regime where the expansion slows down over time, but is still up to three orders of magnitude faster than regular exciton diffusion. This occurs due to phonon-driven scattering between the polariton states within the lightcone and the exciton reservoir. Finally, in the steady-state limit, the system settles into a much slower, exciton-limited diffusive regime. However, the propagation is still enhanced relative to the conventional bare exciton case by the presence of a cavity. Overall, our work provides a material-realistic and predictive theoretical foundation for phonon-mediated polariton transport under technologically relevant conditions.

\section*{Results}

\begin{figure*}[t!]
\includegraphics[width=\textwidth]{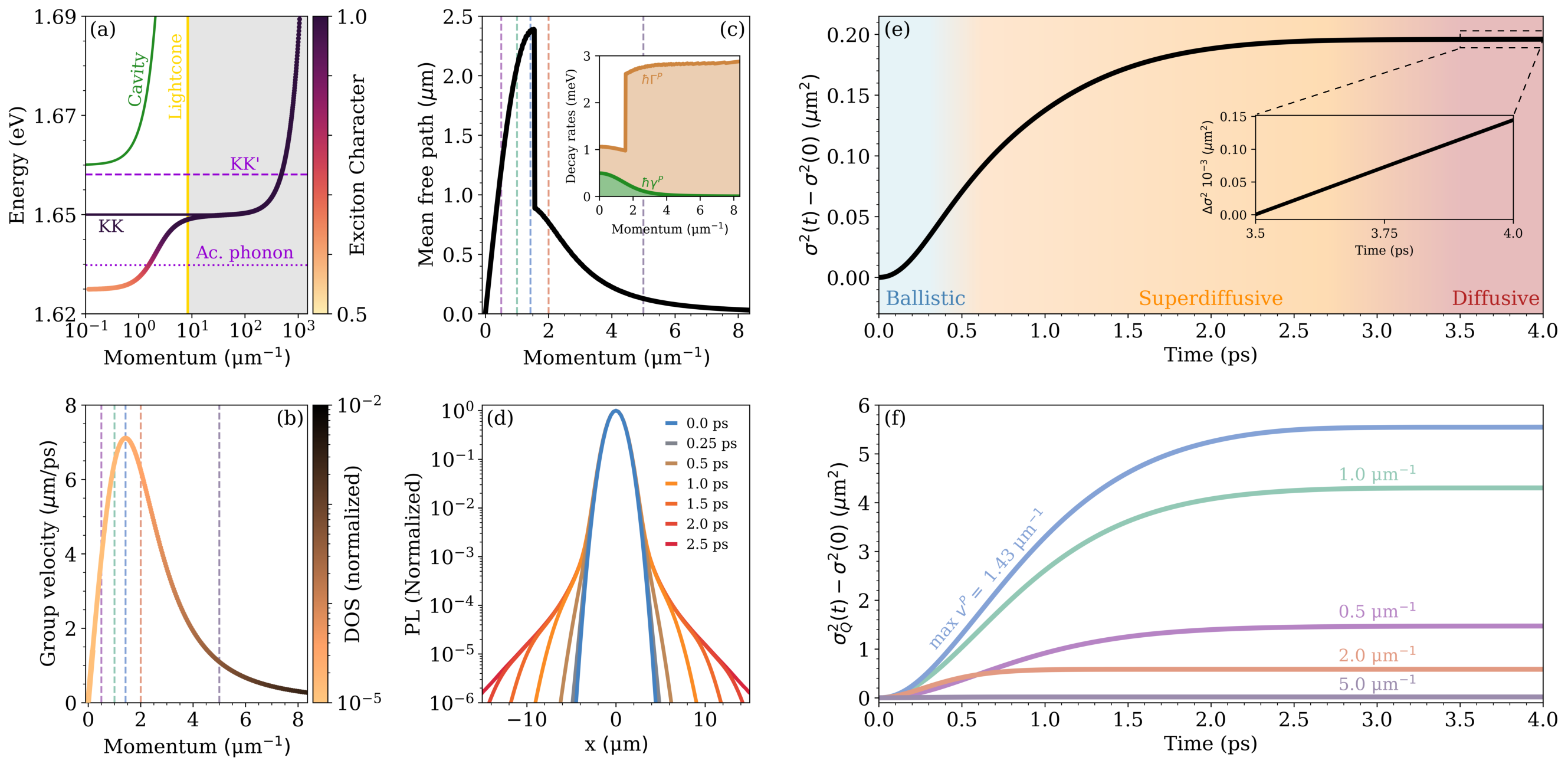}
\caption{(a) Lower polariton energy for a \ce{MoSe2} monolayer integrated within a microcavity that is detuned by $+10$ meV. The colormap shows the excitonic character, which is $62\%$ at $Q=0$. The dashed purple line denotes the position of dark KK$^\prime$ excitons, while the dotted line shows the energy at which intervalley scattering processes into KK$^\prime$ exciton states are possible via zone-edge acoustic phonons. (b) The corresponding group velocity of the lower polariton, with the density of states indicated by the colormap (log scale). (c) The momentum-resolved polariton mean free path with the polariton radiative decay rate (green) and the phonon-induced scattering rate (brown) shown in the inset. (d) Time evolution of the normalized integrated polariton PL (log scale) showing a non-Gaussian expansion. (e) The corresponding broadening of the integrated polariton density, illustrating a crossover from an ultrafast ballistic-like regime in the first few hundred picoseconds (blue shading), to a transient superdiffusive regime from $\sim0.5\rightarrow 3$ ps (orange shading), and finally to a much slower exciton-like diffusion (red shading). The latter regime is characterized by a linear slope, as shown in the inset. (f) The momentum-resolved broadening at five representative momenta (corresponding to the dashed vertical lines in (b) and (c)).  
\label{fig:fig_2}}
\end{figure*}

We consider an hBN-encapsulated \ce{MoSe2} monolayer integrated within a $\lambda/2$ Fabry-P\'{e}rot microcavity that is blue-detuned by $10$ meV at zero in-plane momentum. By solving the Wannier equation \cite{berghauser2014analytical}, monolayer \ce{MoSe2} is found to be a direct semiconductor, where the momentum-dark KK$^\prime$ excitons (dashed purple line in Fig.~\ref{fig:fig_2}(a)) lie about $10$ meV \cite{selig2018dark} above the bottom of the bright KK exciton at $1.65$ eV \cite{ajayi2017approaching}. To account for the coupling between excitons and the cavity mode, a Hopfield transformation is used \cite{Fitzgerald2022,fitzgerald2024circumventing}. The calculated Rabi splitting is $39$ meV, and the lower polariton has an excitonic character of $62\%$ at $Q=0$ (Fig. \ref{fig:fig_2}(a)). Further details on the Wannier-Hopfield model can be found in the Methods section.

In stark contrast to excitons, a key feature of polaritons is their momentum-dependent photonic/excitonic character and sharp dispersion within the lightcone. Focusing on the lower polariton branch, this has two important implications for transport: First, it leads to large group velocities on the order of $1$--$10$ $\mu$m/ps, which peak at the inflection point of the dispersion (Fig.~\ref{fig:fig_2}(b)). These velocities are $\sim 10^{4}$ larger than those of typical excitons in the lightcone ($<1$ nm/ps). This in turn implies very small effective masses \cite{Fitzgerald2022} and density of states (DOS) (colour gradient in Fig.~\ref{fig:fig_2}(b)). Second, it leads to a sharp momentum dependence of the polariton-phonon scattering rate within the lightcone, arising from the opening of specific phonon scattering channels into both momentum-direct (i.e., KK) and momentum-indirect excitons (e.g., KK$^\prime$) \cite{ferreira2022signatures,ferreira2022microscopic,ferreira2024revealing,fitzgerald2024circumventing}. A consequence of this is a non-trivial momentum dependence of the polariton mean free path, $L_Q=v^P_Q\tau_Q$, as shown in Fig.~\ref{fig:fig_2}(c) at $300$ K. At the momentum of $1.55 \ \mu\text{m}^{-1}$, there is a sharp drop in the mean free path, corresponding to the opening of the acoustic phonon scattering channel into the KK$^\prime$ exciton reservoir (purple dotted line in Fig.~\ref{fig:fig_2}(a)). Below this momentum the phonon energy of $18$ meV \cite{jin2014intrinsic} is not sufficient to scatter a polariton into a KK$^\prime$ exciton state \cite{fitzgerald2024circumventing}. These zone-edge phonons are responsible for the ultrafast valley depolarization of carriers observed in  \ce{MoSe2} monolayers \cite{bae2022k}. When this scattering channel opens up, there is a sharp increase in the phonon-induced scattering rate (filled brown area in the inset of Fig.~\ref{fig:fig_2}(c)), $\Gamma_{Q}$, and a corresponding drop in the polariton decay time, $\tau_Q = 1/(2(\gamma_Q+\Gamma_Q))$. Note that at all momenta, both the intravalley KK and intervalley optical channels are open due to the higher phonon energies of $34$ and $33$ meV, respectively \cite{jin2014intrinsic}. This means that lower polaritons can be populated and there is no significant bottleneck \cite{fitzgerald2024circumventing}. The radiative decay rate of the lower polariton $\gamma_Q$ provides a sizeable contribution to the overall decay rate at low momenta within the lightcone (filled green area in the inset of Fig.~\ref{fig:fig_2}(c)). Note that the possible impact of leaky modes at higher angles in the lightcone has been neglected \cite{tassone1997bottleneck}.

\subsection{Distinct polariton transport regimes}
We solve the Boltzmann transport equation to provide a microscopic description of the spatiotemporal dynamics of the exciton polariton Wigner function \cite{hess1996maxwell,rosati2020negative}, $N_{n\vec{Q}}(\vec{r})$, which gives the spatially dependent quasi-distribution of excitons/polaritons with centre-of-mass momentum $\vec{Q}$ and valley/branch $n$. We include exciton/polariton-phonon scattering as well as exciton reservoir states in the KK and KK$^\prime$ valleys. Furthermore, we consider a non-resonant excitation with an initial hot-exciton population in the KK reservoir, i.e., $N_{n\vec{Q}}(\vec{r},t)$ describes an incoherent polariton population. The initial spatial distribution of the exciton polariton cloud corresponds to a Gaussian with a FWHM of $2$ $\mu$m. Further details can be found in the Methods and SI. 

We first focus on the spatiotemporal dynamics of polariton states within the lightcone at room temperature. Figure~\ref{fig:fig_2}(d) shows the time evolution of a slice of the normalized integrated polariton PL (i.e., integrated over all momenta within the lightcone) on a logarithmic scale. A non-Gaussian expansion of the polariton cloud is immediately apparent, with rapid micron-scale broadening occurring in the tail region over about $3$ ps, while the central region of high intensity remains largely unchanged. This is in contrast to bare exciton diffusion, which is typically well modelled by a Gaussian with a time-dependent variance in rotationally invariant systems \cite{kulig2018exciton}. This difference can be attributed to a small population of fast polaritons in the lightcone and a vastly larger population of slow reservoir excitons, which are effectively static on these timescales. As the rapid polaritons radiatively decay or scatter with phonons, the static reservoir is continuously repopulating the polariton occupation, keeping the bulk of the polariton cloud fixed in place. This non-shape-preserving expansion of exciton polaritons has been observed experimentally in TMD \cite{wurdack2021motional}, perovskite \cite{jin2023enhanced}, and amorphous molecular systems \cite{hou2020ultralong}.

Figure~\ref{fig:fig_2}(e) illustrates the time-dependent broadening of the integrated polariton cloud shown in Fig.~\ref{fig:fig_2}(d). We plot the time-dependent contribution of the squared width, i.e., $\sigma^2(t)-\sigma^2(0)$, where $\sigma^2(t)=\int N(\vec{r},t)r^2d^2\textbf{r}/\int N(\vec{r},t)d^2r$ and $\sigma^2(0)$ is determined by the initial spatial profile. Three different regimes of transport are immediately apparent. Below $~300$ fs, there is a fast ballistic-like expansion that is characterized by a quadratic time dependence \cite{rosati2021non} (blue region). In this regime, the polariton occupation grows quickly as it is fed from the non-thermalized KK and KK$^\prime$ exciton reservoirs (see Fig.~S5 in the SI). Consequently, the polariton cloud expands rapidly as any losses are compensated by the efficient in-scattering. As the growth in the polariton occupation saturates and the system approaches a characteristic timescale where the majority of polaritons have scattered with a phonon or radiatively decayed, the rate of expansion drops and the system smoothly evolves into a transient superdiffusive regime \cite{najafi2017super,zhou2022transient} (orange region). Here, the time dependence is no longer quadratic but the polariton cloud still expands at a much larger rate than bare excitons. Finally, as the system tends towards the steady state, the rate of expansion decreases by orders of magnitude and appears flat on the scale of Fig.~\ref{fig:fig_2}(e) (red region). Closer inspection reveals a linear slope (see inset), which is characteristic of the conventional diffusive regime \cite{rosati2021non}. Here, the rate of expansion is about three orders of magnitude smaller than in the earlier transient ballistic-like and superdiffusive regimes. The associated diffusion coefficient corresponds to typical excitonic values of around $1$ cm$^2$/s in  \ce{ MoSe2} monolayers  \cite{wagner2021nonclassical,lo2025transient}. This reveals that the polariton subsystem has reached thermal equilibrium with the exciton reservoir due to the strong scattering with phonons at room temperature. Overall, the time evolution of the polariton broadening shows that ballistic-like propagation of polaritons is possible at room temperature, even for polaritons with a significant excitonic component. However,  this is a short-lived sub-ps behaviour, after which the exciton polariton cloud expands in a cavity-enhanced diffusive fashion.

Further insight is provided in Fig.~\ref{fig:fig_2}(f), where the momentum-resolved broadening is shown for five illustrative momenta indicated by the dashed verticals lines in Figs.~\ref{fig:fig_2}(b) and (c). Because of the one-to-one correspondence between momentum and emission angle, the momentum space distribution of polaritons in the lightcone can be tracked in real space via PL spectroscopy \cite{steger2013long}. Considering the first three momenta (purple, green and blue lines), both the expansion rate and extent of the polariton expansion show a clear dependence on the corresponding group velocity, i.e., a larger $v^P_Q$ leads to a greater broadening. The peak group velocity of $7.1 \ \mu$m/ps at $1.43 \ \mu\text{m}^{-1}$ corresponds to the polariton cloud component that expands most rapidly and furthest (blue line), with $\sigma^2-\sigma(0)^2$ reaching $5.3 \ \mu\text{m}^{2}$ within $2$ ps. For $Q=2$ $\mu \text{m}^{-1}$ (peach curve), we observe a drastic drop in the broadening, despite the large, near-maximum group velocity at this momentum. This is a consequence of the opening of the acoustic phonon scattering channel to KK$^\prime$ excitons at $1.55 \ \mu\text{m}^{-1}$ (Fig.~\ref{fig:fig_2}(c)). This means an increased scattering rate, hence polaritons with these momenta exhibit a shorter ballistic-like and superdiffusive regime, leading to a reduced real-space expansion. Finally, we find for larger momenta, such as $5$ $\mu\text{m}^{-1}$ (lavender curve), a very small broadening that is barely visible on the scale of Fig.~\ref{fig:fig_2}(f). These much slower polaritons possess a very large DOS relative to the more rapid polaritons at lower momenta (Fig.~\ref{fig:fig_2}(b)), hence their contribution to the integrated polariton density in Fig.~\ref{fig:fig_2}(e) is dominant. This explains the significantly reduced broadening ($\sim 25$ times smaller) of the momentum-integrated polariton cloud in Fig.~\ref{fig:fig_2}(e) compared to the momentum-resolved results for the most rapid polaritons shown in Fig.~\ref{fig:fig_2}(f).

%%%%%%%%%%%%%%%%%%%%%%%%%%%%%%%%%%%%%%%%%
% Effective Diffusion coefficient 300 K %
%%%%%%%%%%%%%%%%%%%%%%%%%%%%%%%%%%%%%%%%%
\begin{figure}[t!]
\includegraphics[width=\columnwidth]{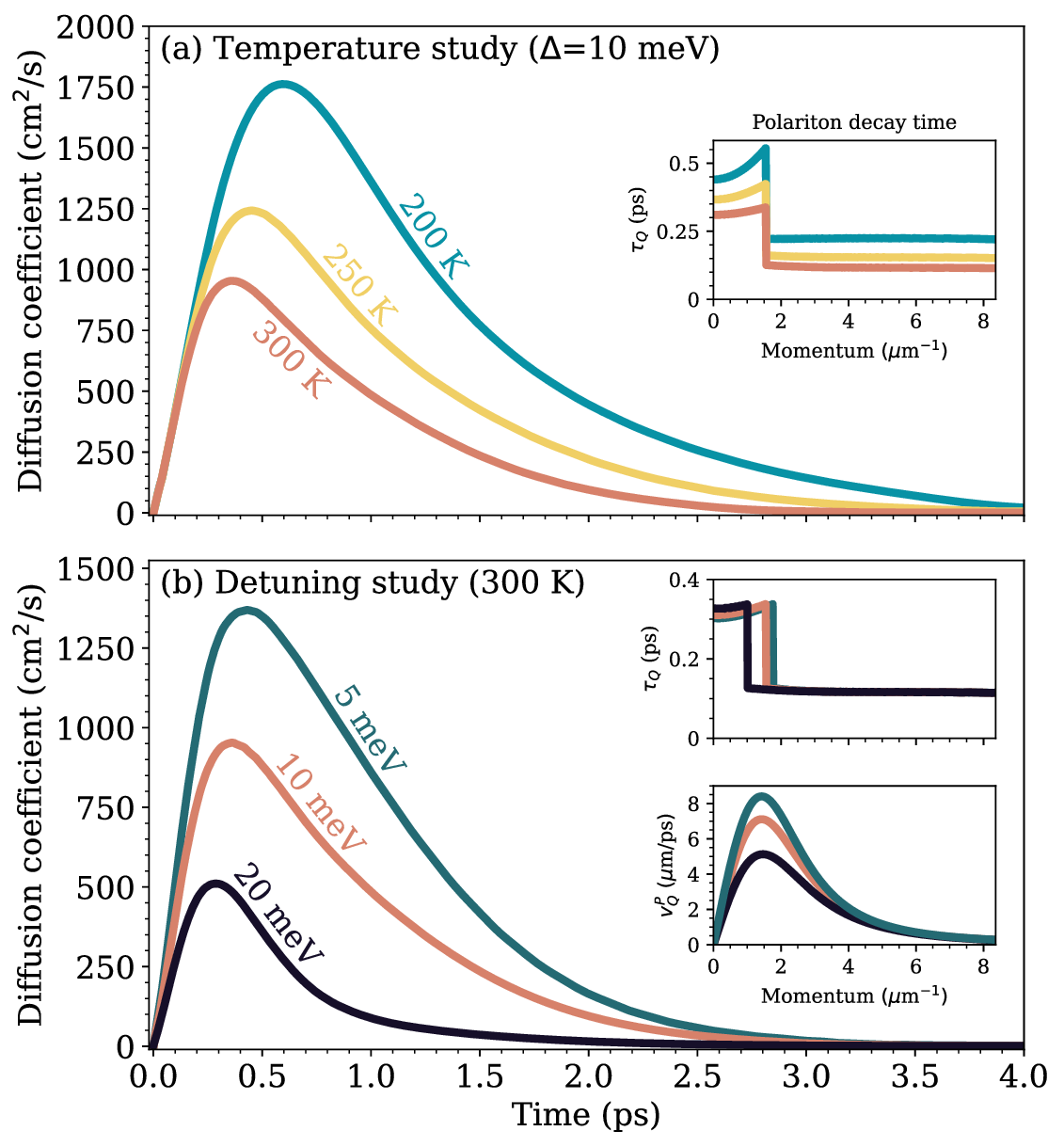}
\caption{(a) Temporal evolution of the temperature-dependent polariton diffusion coefficient at a fixed detuning $\Delta=$+10 meV. The inset shows the corresponding momentum-resolved polariton scattering time, $\tau_Q$. (b) Detuning study of the polariton diffusion coefficient at room temperature. The insets show the momentum-dependent polariton scattering time (top) and the group velocity (bottom). 
\label{fig:fig_3}}
\end{figure}

To further elucidate the different transport regimes, Fig.~\ref{fig:fig_3}(a) shows the effective diffusion coefficient \cite{rosati2020negative, najafi2017super}, $D(t)=\partial_t\sigma^2(t)/4$ of the momentum-integrated polariton cloud for three different temperatures (momentum-resolved results are shown in Fig.~S3). Focusing first on room temperature (red curve), there is an initial linear increase of the diffusion coefficient with time, indicating a ballistic-like regime \cite{rosati2020negative}. After initialization of a hot-exciton population, there is a swift energy relaxation via efficient phonon emission from both the KK and KK$^\prime$ exciton reservoir \cite{fitzgerald2024circumventing}. This leads to a rapid occupation of polariton states within the lightcone over a $100\text{--}300$ fs timescale (see Fig.~S5), and means that any scattering with phonons or radiative decay in that time period is compensated by the feeding of polaritons from the exciton reservoirs. In a sense, the efficient phonon-driven scattering into polariton states mimics a driving term, and hence leads to a ballistic-like transport. 
Next, the diffusion coefficient peaks at a large magnitude of $954$ cm$^2$/s at $0.36$ ps, about three orders of magnitude larger than typical exciton diffusion coefficients \cite{wagner2021nonclassical,lo2025transient}. Over the next $2-3$ ps, the diffusion decreases, but still remains significantly enhanced compared to the cavity-free case. For example, at $2$ ps, the diffusion coefficient is still about $10$ times larger than conventional exciton values. This intermediate superdiffusive regime corresponds to a non-equilibrium state, where the ballistic-like propagation has been terminated by polariton-phonon scattering into the exciton reservoirs, which acts as a frictional mechanism. Eventually, the diffusion limits to a small constant value, indicating the crossover into regular exciton-limited diffusion at steady state described by Fick's law \cite{rosati2020negative}. Our results are consistent with reports of a dynamic crossover from a ballistic to diffusive regime for polaritons with a predominantly excitonic character in molecular materials \cite{balasubrahmaniyam2023enhanced}. Similar transient superdiffusive transport has also been observed for highly hot quasi-free carrier dominated expansion in TMDs \cite{lo2025transient}, hybrid perovskites \cite{guo2017long}, and silicon \cite{najafi2017super}.

%%%%%%%%%%%%%%%%%%%%%%%%%%%%%%%%%%%%
% Temperature dependence diffusion %
%%%%%%%%%%%%%%%%%%%%%%%%%%%%%%%%%%%%

We find a very similar ballistic-like evolution for all three temperatures up to times of about $300\text{--}500$ fs (Fig.~\ref{fig:fig_3}(a)). This is because the in-scattering mechanism is due to phonon emission, which is only weakly sensitive to temperature. We also observe both an increased maximum diffusion coefficient ($1243$ and $1762$ cm$^2$/s for $250$ and $200$ K, respectively) and that it takes a longer time to reach this peak as temperature decreases ($0.45$ and $0.59$ ps for $250$ and $200$ K, respectively). This is due to an enhanced and extended superdiffusive regime at lower temperatures, which is caused by a longer polariton-phonon scattering time within the lightcone (see the inset). For example, at $3$ ps the diffusion coefficient for the system at $200$ K is about $24$ times larger than the corresponding value at $300$ K. This is in contrast to the weak temperature dependence of the bare exciton diffusion coefficient in \ce{MoSe2} monolayers (see Fig.~S8).

%%%%%%%%%%%%%%%%%%%%%%%%%%%%
% Detuning study diffusion %
%%%%%%%%%%%%%%%%%%%%%%%%%%%%

A key advantage of polaritons is the external control of their dispersion via cavity detuning, $\Delta$. In the context of transport, this is especially valuable as it allows for control over both the group velocity and available polariton-phonon scattering channels \cite{ferreira2022signatures,ferreira2024revealing,fitzgerald2024circumventing}. Figure~\ref{fig:fig_3}(b) shows the detuning dependence of the diffusion coefficient at $300$ K. 
For smaller detuning, we find a larger peak diffusion coefficient, reaching $1370$ cm$^2$/s for $\Delta=5$ meV. This value is about $2.7$ times larger than that observed for $\Delta=20$ meV. This enhanced superdiffusivity is a consequence of the larger group velocity (bottom inset), which grows with decreasing detuning as polaritons gain photonic character. In contrast to the group velocity, the scattering rate is relatively unaffected by  detuning, and only the momentum of the opening of the acoustic phonon channel slightly shifts (top inset). As a result,  there is only a small dependence of the peak polariton diffusion on detuning ($0.29$ and $0.42$ ps for $+20$ and $+5$ meV detuning, respectively). However, the less-detuned cavity shows a prolonged superdiffusivity. For example, at $3$ ps the $\Delta=5$ meV system exhibits a diffusion coefficient about $7$ times larger than the $\Delta=20$ meV cavity. These results illustrate that the crossover from ballistic-like to  conventional diffusive regime is sensitive to the excitonic/photonic character of polaritons \cite{balasubrahmaniyam2023enhanced, xu2023ultrafast}. As the cavity is blue-detuned, the lower polariton branch becomes more exciton-like and hence shows a reduced and shorter-lived superdiffusive expansion.

\begin{figure}[t!]
\includegraphics[width=\columnwidth]{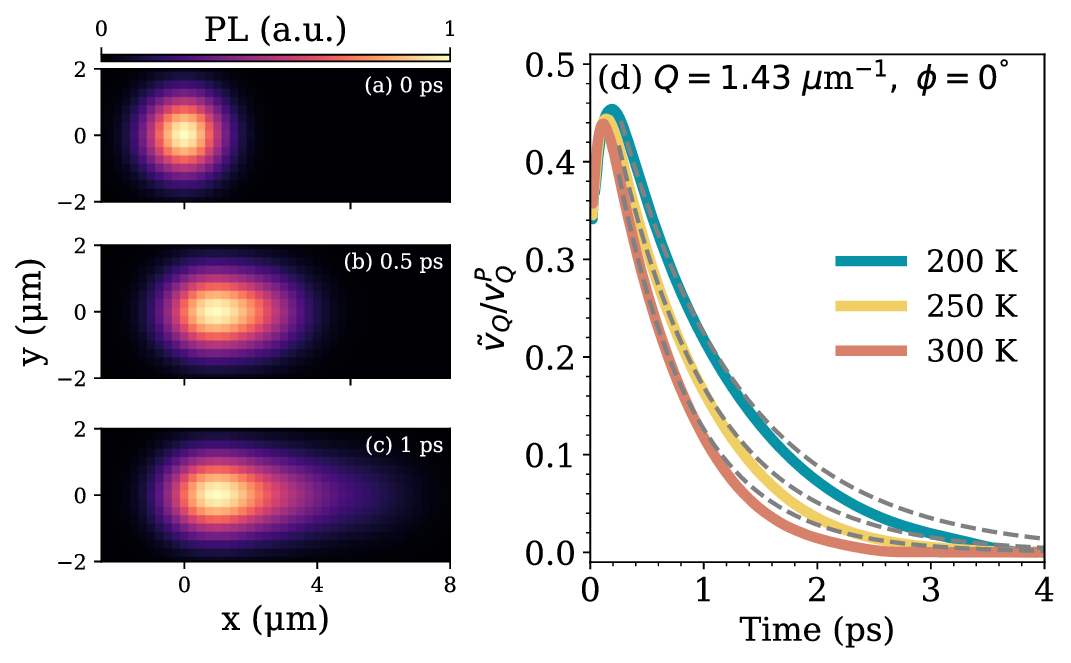}
\caption{Ballistic-like to superdiffusive transport regime. (a)-(c) Photoluminescence maps of the momentum- and angle-resolved polariton density shown at the fixed momentum $Q=1.43 \ \mu\text{m}^{-1}$ (corresponding to the maximum group velocity $v^P_Q$, see Figs.~\ref{fig:fig_2}(b) and (f)) and in-plane angle $\phi=0$ for three fixed times. (d) Ratio of the effective velocity, $\tilde{v}_{Q}=d\braket{x}_Q/dt$, and the group velocity, $v_Q^P$, at three different temperatures. The grey dashed lines show a simple decay model based on the probability that a polariton has scattered with a phonon or radiatively decayed into an external photon. 
\label{fig:fig_4}}
\end{figure}

\subsection{Renormalization of polariton velocity}
Now, we further explore the ballistic-like and the superdiffusive transport regime by studying the momentum- and angle-resolved polariton density, i.e., polariton propagation in a particular in-plane direction. This is an experimentally relevant quantity that is linked to the PL measured at a distance away from the excitation point, which selectively probes polaritons that have propagated in a particular direction \cite{steger2013long,wurdack2021motional,xu2023ultrafast}. Figs.~\ref{fig:fig_4}(a)-(c) show a typical PL density map during the first picosecond of expansion, for an in-plane angle $\phi=0$ and momentum $Q=1.43$ $\mu$m$^{-1}$, corresponding to the maximum group velocity $v^P_Q$ (blue line in Fig.~\ref{fig:fig_2}(f)). It highlights a rapid propagation of angle-resolved polaritons over microns, even at room temperature and in the presence of strong polariton-phonon scattering. 

To quantify directional polariton transport, we introduce the first moment of the exciton polariton cloud, $\braket{x}_Q = \int d^2\textbf{r} N_Q(\textbf{r},t) x /\int d^2\textbf{r} N_Q(\textbf{r},t)$, and take the time derivative to give an effective velocity, $\tilde{v}_{Q}=d\braket{x}_Q/dt$. In the limit of no scattering with phonons, this will simply be the group velocity extracted from the polariton dispersion (see Fig.~S2). In Fig.~\ref{fig:fig_4}(d), the ratio $\tilde{v}_{Q}/v_Q^P$ is shown for three different temperatures at $Q=1.43$ $\mu$m$^{-1}$ (different momenta can be found in Fig.~S4). For all temperatures, the effective velocity is strongly renormalized due to scattering with phonons, even in the sub-ps ballistic-like regime. On an ultrashort timescale of $~100$ fs, we find a small rise of the effective velocity, which is attributed to the non-resonant excitation. In fact, we find that the effective group velocity of polaritons is quite sensitive to the initial conditions of the system (see Fig.~S7). Surprisingly, we observe that the maximum effective velocity is quite temperature-independent reaching a value of about $0.45 \times v_Q^P$ for all three temperatures considered. This is a consequence of the weak temperature dependence of the in-scattering from the initial hot-exciton distribution. 

The decrease in $\tilde{v}_Q$ over time is much less efficient at lower temperatures. For example, at $3$ ps the effective velocity of $0.019\times v_Q^P$ at $200$ K is about $90$ times larger than the respective room temperature value.
This can be understood using a simple model that describes the probability over time of a polariton undergoing a scattering event or radiatively decaying: $x_Q(t)=\int_0^tdt'\ \exp[-(\Gamma_Q + \gamma_Q)t'] v_Q^P $, suggesting the identification of an effective velocity $\tilde{v}_Q(t)\propto \exp[-(\Gamma_Q + \gamma_Q)t']v_Q^P $. We find that this provides a reasonable fit to the results (dashed grey lines in Fig. \ref{fig:fig_4}(d)), especially considering that this model does not account for the generation of new polaritons from the (nearly) static exciton reservoir. This explains why the model underestimates the drop in velocity with time. It does not take into account the generation of polaritons at later times within the original excitation area, which anchors $\braket{x}$ to this region and hence reduces the effective velocity. This can be seen in Fig.~\ref{fig:fig_4}(c), where the central region emits the most intense PL.

\subsection{Polariton-mediated transport of dark excitons} \label{sec:polariton_mediated}
\begin{figure}[t!]
\includegraphics[width=\columnwidth]{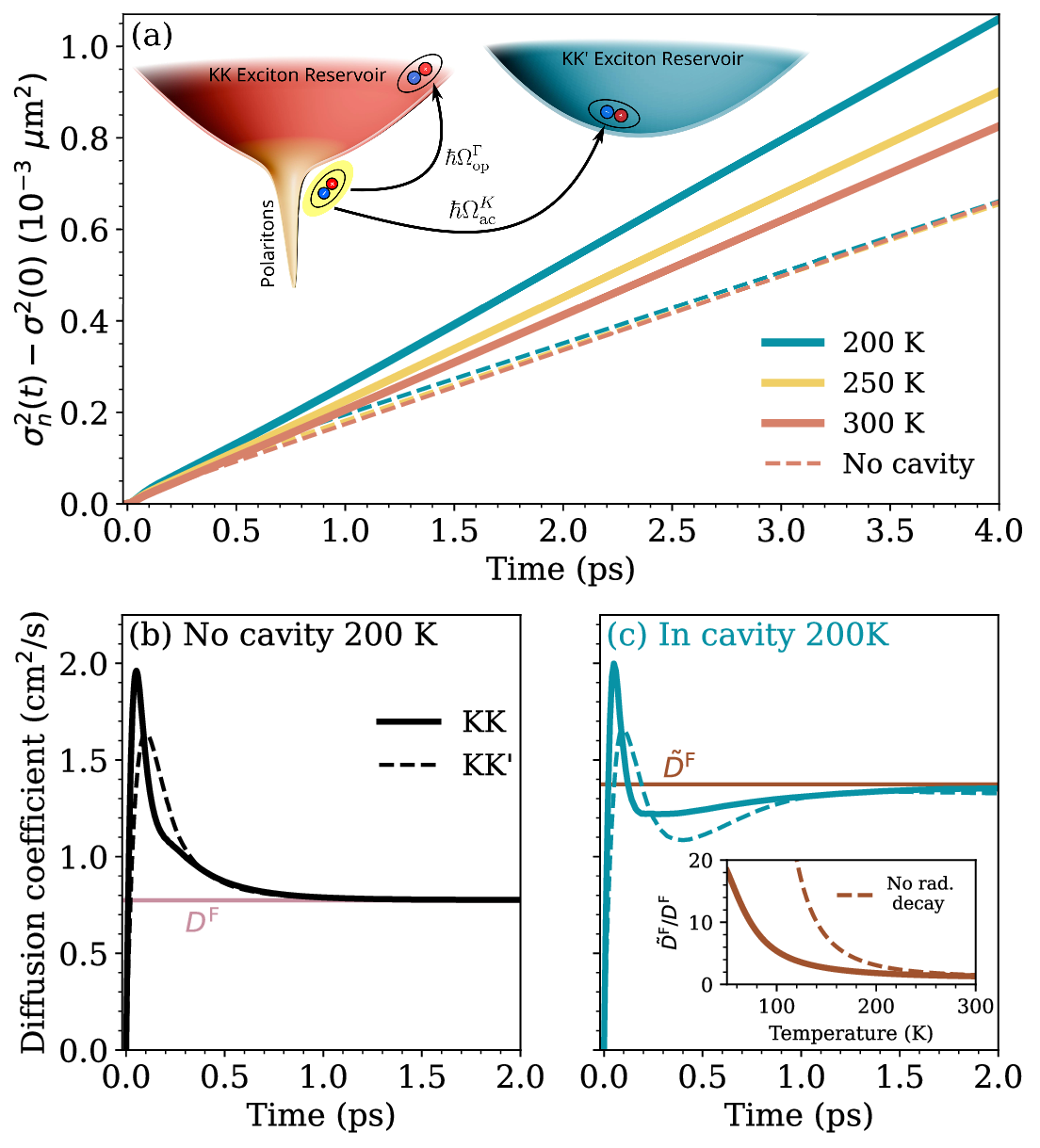}
\caption{Exciton-limited transport regime. (a) Broadening of the total momentum-integrated KK-exciton-polariton density at three different temperatures (including all momenta inside and outside the lightcone). Solid (dashed) lines show the case with (without) a cavity. This reveals an enhanced polariton-mediated exciton propagation in a cavity, as illustrated by the schematic in the inset. (b), (c) The corresponding valley-resolved diffusion coefficients for the system with and without a cavity, respectively. The horizontal lines denote the respective steady-state diffusion coefficient, $D^F$, estimated with Fick's law. The inset illustrates the temperature-dependent cavity enhancement effect, where the dashed line shows the result in the absence of polariton radiative decay.
\label{fig:fig_5}}
\end{figure}

Thus far, we have focused on the propagation of polaritons with momenta within the lightcone. We now consider the total momentum-integrated Wigner function over all states, both inside and outside the lightcone, and focus on the third transport regime: exciton-limited steady-state diffusion. Figure~\ref{fig:fig_5}(a) shows the broadening of the \emph{entire} KK exciton polariton cloud at three different temperatures (equivalent detuning study can be found in Fig.~S6). The corresponding dashed lines illustrate the broadening in a bare exciton system without a cavity. Note that the dashed lines coincide closely, highlighting the relative temperature insensitivity of the bare excitons in contrast to the cavity system (see Fig.~S8). For both systems and all three temperatures, from about $0.5$ ps onward we find that the broadening has an approximately linear time dependence, indicating a rapid thermalization. Here, the transport is dominated by conventional Fickian diffusion \cite{rosati2020negative}. Given the tiny DOS of polaritons within the lightcone, it might be expected that the total broadening would be unchanged in the presence of a cavity. Surprisingly, Fig.~\ref{fig:fig_5}(a) reveals that the expansion significantly speeds up in the presence of a cavity, indicating a cavity enhancement effect in the steady-state \cite{ferreira2022microscopic}. This effect is the strongest at the lowest temperature of $200$ K. For example, at $4$ ps the exciton cloud has expanded $160 \%$ more at $200$ K compared to the cavity-free case, while this is $125\%$ at $300$ K. The mechanism behind this polariton-mediated transport enhancement is illustrated in the inset of Fig.~\ref{fig:fig_5}(a). At elevated temperatures, rapid polaritons in the lightcone can scatter into both the KK and KK$^\prime$ exciton reservoir via phonon absorption. This can happen away from the initial excitation region after a polariton has travelled some distance, leading to an effective transport of the exciton reservoir. While phonon absorption becomes less significant at lower temperatures, the combined effects of increased polariton occupation relative to the exciton reservoir and faster polariton expansion (see Fig.~\ref{fig:fig_3}(a)) lead to a stronger cavity-enhanced expansion.

Turning now to the corresponding effective diffusion coefficients, Figs.~\ref{fig:fig_5}(b) and (c) show the case without and with a cavity at $200$ K, respectively. Here, we plot both the KK (solid) and KK$^\prime$ (dashed) valley-resolved diffusion coefficients, which at thermal equilibrium will limit to the same value due to  efficient phonon-mediated intervalley scattering (see Fig.~S5). Below $0.5$ ps, we find in all cases a sharp transient peak in the diffusion due to the initial hot-exciton population \cite{rosati2020negative} that is only weakly impacted by the presence of a cavity. In contrast, at the steady state the presence of a cavity leads to an increased diffusion for both valleys (for KK this corresponds to the different slopes of the solid and dashed teal lines in Fig.~\ref{fig:fig_5}(a)). For the bare exciton case, the steady state value can be predicted using Fick's law \cite{rosati2020negative,rosati2021non}, $D_n^{\text{F}}=(\hbar/2) \sum_\vec{Q} f_{nQ} v_{nQ}^2\Gamma_{nQ}^{-1}/N_n$, which holds for a quasi-thermalized local distribution where $f_{nQ}$ is the Boltzmann distribution for the valley $n$, and $N_n=\sum_\vec{Q}f_{nQ}$. In a two-valley system, the total steady-state diffusion coefficient is given by the population-weighted valley-resolved coefficients, $D^{\text{F}}=(D_{\text{KK}}^{\text{F}}N_{\text{KK}}+D_{\text{KK}^\prime}^{\text{F}}N_{\text{KK}^\prime})/(N_{\text{KK}}+N_{\text{KK}^\prime})$. This is shown in Fig.~\ref{fig:fig_5}(b) with the horizontal pink line, where $D^{\text{F}}=0.77$ cm$^2$/s, which correctly gives the steady-state limit of the full Boltzmann transport dynamics. Fick's law can also be applied to polaritons \cite{ferreira2022microscopic}, but it is important to correct for the bottleneck effect \cite{fitzgerald2024circumventing}. In the limit of not too strong radiative decay, we can take a corrected Fick's law, $\tilde{D}_{\text{KK}}^{\text{F}}$, with $\Gamma_{nQ} \rightarrow\Gamma_{nQ}+\gamma_{nQ}$, and $\tilde{f}_{nQ}=\Gamma_{nQ}/(\gamma_{nQ}+\Gamma_{nQ})f_{nQ}$ as the modified occupation factor (see SI for further details). Because of the contribution of the high-velocity polaritons, we find that $\tilde{D}^{\text{F}}$ has a much higher value of $1.37$ cm$^2$/s compared to $D^{\text{F}}$, shown by the brown horizontal line in Fig.~\ref{fig:fig_5}(c). This compares to $2.32$ cm$^2$/s, if the polariton occupation is incorrectly assumed to follow a Boltzmann distribution. Figure~\ref{fig:fig_5}(c) reveals that the full dynamics of both the KK and KK' exciton reservoirs correctly limit towards the polariton-modified Fick's law as the system thermalizes, exhibiting a cavity-induced enhancement of $\tilde{D}^{\text{F}}/D^{\text{F}}=1.77$. Given the excellent agreement between the polaritonic Fick's law and the steady-state limit of the numerical results, we can explore the cavity enhancement over a large temperature range at greatly reduced computational cost. The inset of Fig.~\ref{fig:fig_5}(c) reveals that at lower temperatures the cavity enhancement of the diffusion coefficient grows rapidly, reaching $18.3$ at $50$ K. This stems from a higher polariton occupation relative to the reservoir and an increased mean free path of the excitons/polaritons. In the absence of radiative decay (dashed line), there is no polariton bottleneck and an even larger diffusion enhancement is possible, reaching $\sim 10^4$ at $50$ K \cite{ferreira2022microscopic}. While the impact of the exciton reservoir on spatiotemporal polariton dynamics is established in the literature \cite{wertz2010spontaneous,fitzgerald2024circumventing}, we show here that also a polariton population measurably influences the exciton reservoir, despite its small density of states.

\section*{Discussion}
In this work, we provide new microscopic insights into ultrafast transport phenomena of exciton-polaritons in 2D semiconductors, crucially taking into account the full exciton energy landscape. We find three distinct transport regimes, ranging from the initial sub-ps ballistic-like regime with phonon-induced renormalization of the effective polariton velocity, to a few-ps superdiffusive regime with a massive polariton-enhanced diffusion, and a slower steady-state conventional exciton diffusion. For the latter, we predict a polariton-mediated enhancement of exciton reservoir diffusion via phonon-driven re-population of higher energy dark excitons from highly mobile polaritons. The results highlight the importance of polariton-phonon scattering at ambient temperatures, as well as of the crucial role of the entire exciton reservoir as scattering partners for polaritons. The developed approach can be applied to a broader class of excitonic 2D materials, such as metal-halide perovskites \cite{konig2025magneto}, as well as to spatially structured cavities that host, for example, trapped \cite{wurdack2021motional} and lattice \cite{jacqmin2014direct} polaritons.

\section*{Methods}
Using a material-specific Wannier–Hopfield method \cite{Fitzgerald2022}, we calculate the dynamics of the space- and momentum-resolved polariton Wigner function \cite{hess1996maxwell}, $N_{n\vec{Q}}(\vec{R},t)$, using the Heisenberg equation of motion \cite{kira2006many}. Here, $n$ denotes both the polariton branch and the exciton valley, while $\vec{Q}$ is the centre-of-mass momentum and $\vec{R}$ the spatial coordinate. We restrict our focus to the low-excitation regime, where the dominant source of scattering is exciton/polariton-phonon scattering. In analogy to the bare exciton case \cite{rosati2020negative,rosati2020strain,rosati2021non}, the resulting spatiotemporal dynamics is described by a Boltzmann transport equation (see SI for further details)
\begin{align}
    \partial_t N_{n\vec{Q}}(\vec{R},t) &= (-\vec{v}_{n\vec{Q}}\cdot \nabla-2\Gamma_{n\vec{Q}}-2\gamma_{n\vec{Q}})N_{n\vec{Q}}(\vec{R},t) \nonumber
    \\ &+ \sum_{m\vec{Q}'} \Gamma_{m\vec{Q}',n\vec{Q}}  N_{m\vec{Q}'}(\vec{R},t) \label{eq:boltzmann}
\end{align}
where $v_{n\vec{Q}}=\partial_Q E_{n\vec{Q}}/\hbar$ is the group velocity, $\gamma_{n\vec{Q}}$  the radiative decay rate, $\Gamma_{n\vec{Q}}$  the exciton/polariton-phonon dephasing rate, and $\Gamma_{m\vec{Q}',n\vec{Q}}$  the in-scattering rate from the state $\ket{m,\vec{Q}'}$ into $\ket{n,\vec{Q}}$ via emission or absorption of phonons. The first term in Eq.~(\ref{eq:boltzmann}) describes ballistic motion, where each state moves unimpeded through space with the velocity $\vec{v}_\vec{Q}$. Exciton/polariton-phonon scattering is treated on the level of the second-order Born-Markov approximation \cite{ferreira2022microscopic,ferreira2022signatures,ferreira2024revealing,fitzgerald2024circumventing}. For polaritons, this is dominated by scattering into exciton states outside the lightcone.

The exciton-microcavity coupling is modelled  by taking into account penetration of the cavity field into the DBR mirrors (see SI for details). The latter are taken to be $8$ periods of \ce{SiO2} and \ce{NbO2}, giving a reflectance of $98.7\%$ at the centre of the stopband. Based on calculations of the spatially homogeneous dynamics \cite{fitzgerald2024circumventing}, the upper polariton is found to have a negligible contribution and is hence ignored. To describe non-resonant excitation of the system, all dynamical calculations are initialized with a Gaussian distribution in the KK exciton reservoir, centred at $50$ meV and with a width of $1$ meV. To mimic the excitation of the system by a laser with a typical finite beam waist, the spatial distribution of the polariton cloud is initialized as a Gaussian with a FWHM of $2$ $\mu$m. The results are found not to be sensitive to the choice of this width. Because of the system's translational invariance, the exciton-polariton cloud remains rotationally invariant at all times.

\begin{acknowledgments}
The authors acknowledge funding from the Deutsche Forschungsgemeinschaft (DFG) via the regular project 524612380. Calculations for this research were conducted on the Lichtenberg high-performance computer of the TU Darmstadt (Project 2373).
\end{acknowledgments}

%\bibliography{bib}

%\clearpage
%\newpage
\onecolumngrid

\setcounter{equation}{0}
\setcounter{figure}{0}
\setcounter{section}{0}
\setcounter{subsection}{0}

\renewcommand{\thesection}{S.\Roman{section}}
\renewcommand{\theequation}{S.\arabic{equation}}
\renewcommand{\thefigure}{S\arabic{figure}} % Set figure label

\section*{Supplementary materials}

\subsection{Methods}

\subsubsection{Modelling excitons with the Wannier equation}
Exciton energies, wavefunctions, and oscillator strengths are calculated using the Wannier equation \cite{berghauser2014analytical,selig2016excitonic}, with density functional theory (DFT) input used to characterise the two-band parabolic approximation for the electronic bandstructure \cite{kormanyos2015k}. The Wannier equation is solved for the bright KK, the momentum-dark KK’, and K$\Lambda$ 1s excitons. The latter are found to be approximately $140$ meV higher in energy than the KK excitons, and hence have a negligible contribution to the dynamics explored in this work. The spectral position of KK excitons in hBN-encapsulated  \ce{MoSe2} monolayers is fixed to $1.65$ eV to match PL measurements \cite{ajayi2017approaching}. The screened Coulomb potential is modelled as a generalised Keldysh potential \cite{brem2019intrinsic} using DFT-derived dielectric constants for the TMD monolayer ($\epsilon_\perp=7.2$ and $\epsilon_\parallel = 16.8$) \cite{laturia2018dielectric}, and $\epsilon_{\text{sub}}=4.5$ for the surrounding hBN layers. 

To estimate the optical matrix element, which dictates the strength of the exciton-photon coupling and the Rabi splitting in a cavity, a two-band $\vec{k}\cdot\vec{p}$ expansion is used \cite{haug2009quantum}, $M_0 =m_0\sqrt{E_g/(4m_r)}$. Here, $m_r = m_em_h/(m_e + m_h)$ is the reduced electron-hole mass, $m_0$ is the free electron mass, and $m_{e/h}$ the effective electron/hole mass. The exciton radiative lifetime is given by \cite{kira2006many}
\begin{equation}
    \gamma^X = \frac{e^2M_0^2|\psi^X(\vec{r}=0)|^2}{2m_0^2\epsilon_0 cE^X_0},
\end{equation}
where $E^X_Q$ and $\psi^X$ are the 1s KK exciton energy and wavefunction, respectively.

\subsection{Exciton-photon coupling in a cavity}
\begin{figure}[t!]
\includegraphics[width=\columnwidth*3/5]{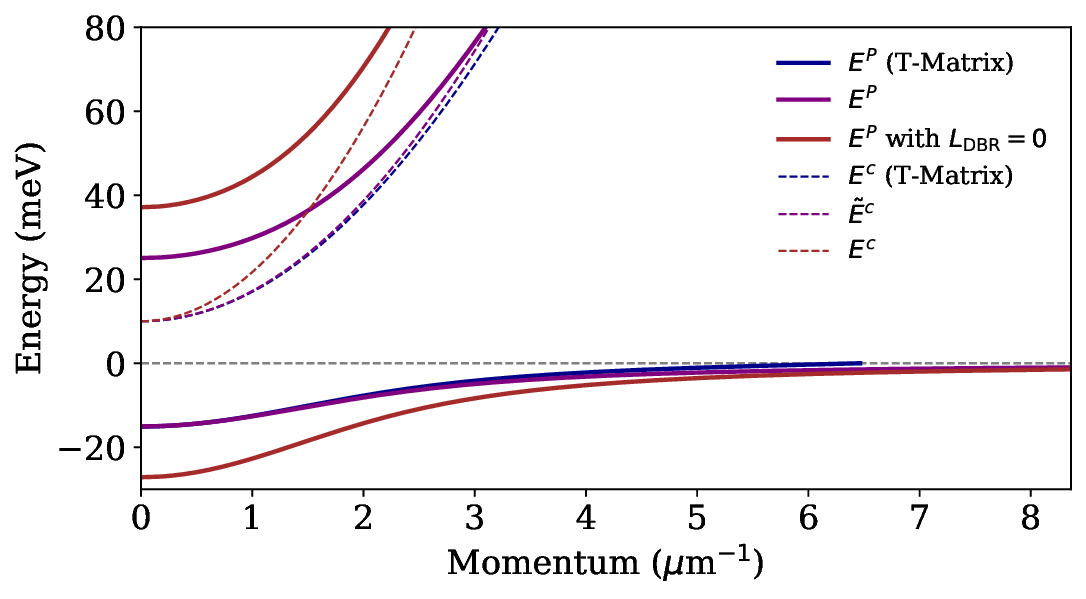}
\caption{Energy (relative to the 1s KK exciton) of the lower and upper polaritons (solid lines), and cavity mode energies (dashed lines) for a $10$ meV blue detuned $\lambda/2$ cavity with a MoSe2 monolayer in the centre. The end mirrors are two identical 8-period \ce{NbO2}/\ce{SiO2} DBRs with a stop-band centre reflectance of $98.7\%$. The dashed and solid brown lines show the cavity, $E^c$, and polariton energy, respectively, within the perfect-mirror model (no phase penetration into the DBR, $L_{\text{DBR}}=0$). Likewise, $\tilde{E}^c$ (dashed purple curve) and $E^P$ (solid purple curve) give the cavity and polariton energy, respectively, including corrections for the DBR phase penetration. Excellent agreement between this model and the T-matrix method (blue curves) is found. Note that the T-matrix fit for the lower polariton energy is terminated at the start of the leaky-mode region \cite{tassone1997bottleneck}, where the single optical mode assumption breaks down.
\label{fig:fig_S1}}
\end{figure}

We focus on the lowest energy optical mode of a symmetric $\lambda/2$ Fabry-P\'{e}rot microcavity in vacuum, where a single-mode approximation is valid due to the large free-spectral range. For small microcavities constructed from distributed Bragg reflectors (DBRs), it is essential to account for the phase penetration of the cavity field into the mirrors to accurately calculate the cavity mode dispersion and the exciton-cavity coupling. For frequencies close to the stop-band centre frequency, $\omega_s$, the reflection coefficient has a linear dependence on phase, leading to a simple expression for the penetration depth \cite{koks2021microcavity}
\begin{equation}
    L_{\text{DBR}}=\frac{\lambda_s/2}{n_H-n_L},
\end{equation}
where $\lambda_s=2\pi c/\omega_s$, and $n_{L/H}$ are refractive indices of the alternating low/high-index layers of the DBR. This expression is valid for a DBR where the first interface is the high-index material. For the alternative situation, where the first interface is the low-index material, the penetration depth is given by $(n_Ln_H\lambda_s/2)/(n_H-n_L)$ \cite{koks2021microcavity}. Assuming frequencies close to the stop-band, the modified resonance condition of the cavity can be written as \cite{panzarini1999cavity}
\begin{equation}
   \tilde{E}^c(\theta) = \frac{L_c E^c(\theta) + L_{\text{DBR}}\hbar\omega_s(\theta)}{L_c + L_{\text{DBR}} },
\end{equation}
where $\theta$ is the angle of incidence, $L_c$ is the physical cavity length, and $E^c(\theta)=\hbar \pi c/(L_c \cos(\theta))$ is the Fabry-P\'{e}rot mode energy in the absence of any phase delay. The angle $\theta$ also corresponds to the photon propagation angle within the cavity, as we assume a vacuum for both the external and internal media. A simple expression for the angle-dependent stop-band frequency can be derived by considering the phase change across one period of the DBR
\begin{equation}
    \omega_s(\theta) = \frac{2 \omega_s(0)}{\cos(\theta_L) + \cos(\theta_H)},
\end{equation}
where $\theta_{L/H}$ are the angles of propagation in the low/high-index layers, which can be determined from Snell's law. The angle can linked to the momentum of a particle via the relation $Q=\sin(\theta)E(\theta)/(\hbar c)$, where $E$ is the particle's dispersion. Figure~\ref{fig:fig_S1} shows a comparison of the model (dashed purple curve) against the commonly used perfect-mirror model ($L_{\text{DBR}}=0$, dashed brown curve), and the exact cavity mode dispersion calculated with the T-matrix method \cite{konig2025magneto} (dashed blue curve). Excellent agreement is found at small angles/momenta.

The cavity decay rate is estimated as \cite{Fitzgerald2022} 
\begin{equation}
    \hbar \kappa = \frac{\hbar c (1-R_s)}{2(L_c + L_{\text{DBR}})}
\end{equation}
with \cite{panzarini1999cavity} $R_s\approx 1 - 4\left(\frac{n_L}{n_H}\right)^{2 N}$, where $N$ is the number of periods in the DBR. The exciton-cavity coupling depends on the polarization for non-zero angles. In this work we focus exclusively on TM polarization, yielding \cite{panzarini1999cavity,Fitzgerald2022}
\begin{equation}
    g \approx \sqrt{ \frac{2 c \hbar \gamma^X}{L_c+L_{\text{DBR}}}},
\end{equation}
which is angle/momentum independent. For the 8-period \ce{NbO2}/\ce{SiO2} DBR considered in this work, we find a Rabi splitting of $2g=39$ meV, in good agreement with experimental values from similar microcavity systems \cite{dufferwiel2015exciton}. This compares to a value of $63.5$ meV using the perfect-mirror model. As seen for the lower polariton in Fig.~\ref{fig:fig_S1}, the corrected model (purple curve) gives a much better agreement with the extracted T-matrix dispersion (blue curve) when compared to the perfect-mirror model (brown curve).

An advantage of the model described here is that all quantities can be calculated analytically from the DBR parameters, eliminating the need for full-wave simulation of Maxwell's equations, such as the T-matrix method \cite{konig2025magneto}. Furthermore, it is simple to utilize the Hopfield method (i.e., a coupled oscillator model) to describe the strong coupling between the bright KK exciton and the cavity photon. The polariton energies and eigenfunctions (Hopfield coefficients) can be expressed analytically for a one-exciton/one-photon system. For the lower polariton branch we find \cite{haug2009quantum}
\begin{align}
    E_{\vec{Q}}^{P} 
    &= \frac{E^{c}_{\vec{Q}}+E^{X}_{\vec{Q}}}{2}-\frac{1}{2}\sqrt{(E^{c}_{\vec{Q}}-E^{X}_{\vec{Q}})^2+4g^2}
    \\
    U_{\vec{Q}} &= \sqrt{ \frac{E_{\vec{Q}}^{P} -E^{X}_{\vec{Q}}} {2E_{\vec{Q}}^{P} -E^{X}_{\vec{Q}}  -E^{c}_{\vec{Q}} }}
    \\
    V_{\vec{Q}} &= \sqrt{ \frac{E_{\vec{Q}}^{P} -E^{c}_{\vec{Q}}} {2E_{\vec{Q}}^{P} -E^{X}_{\vec{Q}}  -E^{c}_{\vec{Q}} }}
    ,
\end{align}
where $U_\vec{Q}/V_{\vec{Q}}$, give the photonic/excitonic character of the exciton polariton. The resulting polariton radiative decay rate is given by the bare cavity decay rate scaled by the photonic Hopfield coefficient (see the green shaded curve in the inset of Fig.~2(c) in the main text)
\begin{equation}
    \hbar \gamma^P_\vec{Q} = |U_\vec{Q}|^2 \hbar \kappa. \label{eq:polariton_decay}
\end{equation}
In principle, the coupling and hence Hopfield coefficients can be extracted from a T-matrix simulation, giving exact results at the level of Maxwell's equations \cite{konig2025magneto}. In practice, the issue with this approach is that other photonic modes (in particular, leaky modes of the DBR) can interact with the exciton, disrupting the fitting procedure at large angles. This could be overcome by including additional photonic modes in the Hopfield method, but the computational cost would rise dramatically. Because the model neglects leaky modes, it fails to account for large-angle leakage of the Bragg mirrors \cite{tassone1997bottleneck}, thereby underestimating of the polariton radiative decay at high momenta within the lightcone.

\subsubsection{Exciton- and polariton-phonon coupling} \label{sec:phonon_scattering}
Exciton- and polariton-phonon scattering is taken into account by considering acoustic (longitudinal and transverse) and optical (longitudinal, transverse, and the out-of-plane A$_1$) phonons at the high-symmetry points of the phonon dispersion \cite{jin2014intrinsic}. Phonons at these momenta are relevant for phonon-driven intra- and intervalley exciton scattering \cite{fitzgerald2024circumventing}. For simplicity and to ease the computational burden, the set of phonons at each high-symmetry point is treated as one averaged acoustic and one averaged optical phonon \cite{lengers2021phonon}. The relevant phonon modes for this work are the $\Gamma$ phonons, which drive intravalley scattering, and $K$ phonons, which can scatter conduction electrons between the K and K$^\prime$ valley of the electronic bandstructure to create KK$^\prime$ excitons. In particular, coherent phonon studies have shown that longitudinal acoustic K phonons are responsible for the strong valley depolarization observed in \ce{MoSe2} monolayers \cite{bae2022k}. The long-range $\Gamma$ acoustic phonons are approximated with a linear dispersion (Debye model), while all other phonons are treated with a constant energy approximation (Einstein model). 

Electron-phonon scattering matrix elements, $D_{\vec{q}}^{c/v}$, are determined within the deformation potential approximation, with relevant parameters taken from DFT studies \cite{jin2014intrinsic}. Solving the Wannier equation and converting to the exciton basis then gives the exciton-phonon matrix elements \cite{selig2016excitonic,brem2019intrinsic,selig2018dark}. The relative phases of the electron-phonon matrix elements are not specified in Ref.~\citenum{jin2014intrinsic}, but they are important for determining the strength of the intravalley exciton-phonon coupling, which depends on the difference of the electron and hole coupling to phonons. They must be inferred from the dominant physical scattering mechanism for each phonon species \cite{lengers2020theory}. Considering acoustic modes, we assume that the coupling strength predominantly arises from a non-polar lattice deformation, meaning that the conduction and valence band shift in opposite directions under strain ($|D_{\vec{q}}^{c}+D_{\vec{q}}^{v}|$). In contrast, optical modes are assumed to predominantly couple to electrons via the Fr\"{o}hlich interaction, where the two bands shift in the same direction energetically ($|D_{\vec{q}}^{c}-D_{\vec{q}}^{v}|$) \cite{sohier2016two,lengers2020theory}. While the averaging over phonon modes employed in this work leads to a simpler picture of exciton-phonon interactions, it risks inaccurately estimating the exciton-phonon coupling if there is a mix of different mechanisms. In particular, the intravalley scattering of excitons via $\Gamma$ optical phonons is sensitive to the choice of phase. 

After performing a Hopfield transformation, the polariton-phonon scattering strength is given by the exciton-phonon matrix element, $D_{\alpha,\mu \nu ,q }$, scaled by the excitonic component of both the initial and final polariton state: \cite{lengers2021phonon,ferreira2022microscopic,ferreira2022signatures,ferreira2024revealing,fitzgerald2024circumventing} $\tilde{D}_{\alpha,n\vec{Q},m\vec{Q}'}=\sum_{\mu\nu} V_{\mu n,\vec{Q}}^* D_{\alpha,\mu \nu ,|\vec{Q}'-\vec{Q}| } V_{\nu m,\vec{Q}'} $. Here, $q =|\vec{Q}-\vec{Q}'|$ is the momentum imparted by the phonon, and $\alpha$ is the phonon species index. These matrix elements describe both the strength of intravalley scattering within the polariton branches (including the dark reservoir excitons outside the lightcone), and scattering between dark intervalley excitons and polaritons.

\subsection{Polariton Boltzmann transport equation}

Now, we investigate polariton dynamics in a spatially homogeneous system that has an initial inhomogeneous spatial distribution, such as a polariton gas generated by a laser with a finite beam waist. We begin by introducing a polaritonic density matrix, $\rho_{n\vec{Q},m\vec{Q}'}=\braket{\hat{P}_{n\vec{Q}}\hat{P}_{m\vec{Q}'}}$, 
which contains information about the spatial distribution of polaritons in the off-diagonal and their occupation in the diagonals. In direct analogy with the bare exciton case \cite{rosati2020negative,rosati2021non}, we integrate out relative coordinates of the polariton/exciton and define a polaritonic Wigner function for the $n$th branch/valley
\begin{equation}
    N_{n\vec{Q}}(\vec{R})=\sum_{\vec{q}}\braket{\hat{P}_{n\vec{Q}+\vec{q}}^{\dagger}\hat{P}_{n\vec{Q}}}\exp[i\vec{q}\cdot\vec{R}],
\end{equation}
which gives the spatially dependent quasi-probability distribution for a polariton.

To derive the Boltzmann transport equation for the polariton Wigner function, we start by considering the semi-classical dynamics of a particle at a spatial region $\{\vec{R},\vec{Q}\}$, and in a band $E_{n\vec{Q}}$ over an infinitesimal time $dt$ \cite{grosso2013solid}. In the absence of external forces, the coordinates evolve as $\{\vec{R},\vec{Q}\}\rightarrow\{\vec{R}+\vec{v}dt,\vec{Q}\}$, where $\vec{v}_{n\vec{Q}}= \partial_\vec{Q} E_{n\vec{Q}}/\hbar$ is the group velocity. Collision processes (in this case via lattice vibrations) and radiative decay result in a net rate of change in the number of particles in the phase space region $d\vec{R}d\vec{Q}$. Using the Liouville theorem, a condition of detailed balance can be expressed for each polariton/exciton branch/valley
\begin{align}
    N_{n\vec{Q}}(\vec{R}+\vec{v}dt,t+dt)&=N_{n\vec{Q}}(\vec{R},t)+\partial_{t} N_{n\vec{Q}}(\vec{R})|_{\text{scat}}dt +\partial_{t} N_{n\vec{Q}}(\vec{R})|_{\text{rad}}dt \nonumber
    \\
    \therefore \frac{\partial}{\partial t} N_{n\vec{Q}}(\vec{R},t)+\vec{v}_{n\vec{Q}}\cdot\nabla  N_{n\vec{Q}}(\vec{R},t)&=\partial_{t} N_{n\vec{Q}}(\vec{R},t)|_{\text{scat}} + \partial_{t} N_{n\vec{Q}}(\vec{R},t)|_{\text{rad}}, 
\end{align}
where in the second line we expand to first order \cite{grosso2013solid}. The second term on the left describes ballistic propagation (free evolution) of polaritons, where each state moves along the direction of $\vec{Q}$ at the group velocity $\vec{v}_{nQ}$. Using an equation of motion approach \cite{kira2006many, Koenig2016} and assuming a low polariton/exciton density, the scattering terms, $\partial_{t} N_{n\vec{Q}}(\vec{R},t)|_{\text{scat}}$ and $\partial_{t} N_{n\vec{Q}}(\vec{R},t)|_{\text{rad}}$, can be derived by considering the coupling of polaritons to a thermalized phonon bath and external photon ports (quasi-mode approximation), respectively. This procedure is described in detail in Ref.~\citenum{fitzgerald2024circumventing}, so we quote only the final result here (equation (1) of the main text) 
\begin{equation}
    \frac{\partial}{\partial t}N_{n\vec{Q}}(\vec{R},t)= (-\vec{v}_{n\vec{Q}}\cdot \nabla -2\Gamma_{n\vec{Q}}-2\gamma_{n\vec{Q}})N_{n\vec{Q}}(\vec{R},t) + \sum_{m\vec{Q}'} \Gamma_{n\vec{Q},m\vec{Q}'}  N_{m\vec{Q}'}(\vec{R},t). \label{eq:boltzmann}
\end{equation}
For a high-quality symmetric cavity, the polariton decay rate is given by Eq.~\ref{eq:polariton_decay}, i.e., the total cavity decay rate scaled by the photonic Hopfield coefficient \cite{fitzgerald2024circumventing}. The exciton/polariton-phonon scattering rate is calculated on the level of the Born-Markov approximation, describing the probability for a phonon to scatter an exciton/polariton from state $\ket{n\vec{Q}}$ to $\ket{m\vec{Q}'}$. It is given by: 
\begin{equation}
    \Gamma_{n\vec{Q},m\vec{Q}'}=\frac{2\pi}{\hbar}\sum_{\alpha,\pm}
    \left|\tilde{D}_{\alpha,n\vec{Q},m\vec{Q}'} \right|^{2}
\left(\frac{1}{2}\pm\frac{1}{2}+n_{\alpha,|\vec{Q}-\vec{Q}'|}^{\text{phn}}\right) 
 \delta\left(E_{m\vec{Q}'}-E_{n\vec{Q}} \pm E^{\text{phn}}_{\alpha,|\vec{Q}-\vec{Q}'|}\right). 
\end{equation}
Here, the phonon occupation factor, $n_{\alpha \vec{q}}^{\text{phn}}$, is given by a Bose Einstein distribution (phonon bath approximation), $\pm$ denotes phonon emission/absorption, $E^{\text{phn}}_{\alpha,q}$ is the phonon energy, and the delta function enforces strict energy and momentum conservation. The corresponding out-scattering term is given by summing over all possible energy- and momentum-conserving channels: $\Gamma_{n\vec{Q}}=\sum_{m\vec{Q}'}\Gamma_{n\vec{Q},m\vec{Q}'}/2$. This leads to openings of scattering channels at certain momenta of the polariton dispersion (see Fig.~2(c) in the main text). Neglected higher-order dephasing processes are expected to broaden the opening of scattering channels \cite{schilp1994electron}. It is the exciton/polariton-phonon scattering terms of Eq.~\ref{eq:boltzmann} that drive diffusion in the system. They redistribute the polariton occupation primarily toward lower energy states (energy-relaxation) and different orientations in reciprocal space (momentum-relaxation) \cite{rosati2021non}.

Solving Eq.~\ref{eq:boltzmann} poses a challenging numerical problem due to the large range of momentum and spatial scales that must be resolved, spanning from photonic phenomena on the scale of micrometers to excitonic phenomena at the nanometre scale. For the discretisation in momentum space, we use the same method as introduced in Ref.~\citenum{fitzgerald2024circumventing}. We find that using a logarithmic momentum grid works well at accurately describing the dynamics of both polaritons within the lightcone and the exciton reservoir outside. Particle conservation is not enforced and can be used as a check of the numerical calculation. In principle, polariton population is lost due to radiative recombination within the lightcone, but these states represent a tiny fraction of the total KK exciton reservoir, meaning that the total polariton number is approximately conserved over the timescales we explore. We solve the spatiotemporal dynamics on a two-dimensional Cartesian grid using a second-order centred finite-difference approximation. To mimic non-resonant excitation, all dynamics (unless otherwise stated) are initialized with a Gaussian distribution of width $1$ meV and centred at $50$ meV in the KK reservoir. A study of the impact of the initialization energy on the results is presented in Section~\ref{sec:init}.

\begin{figure}[t!]
\includegraphics[width=\columnwidth*3/5]{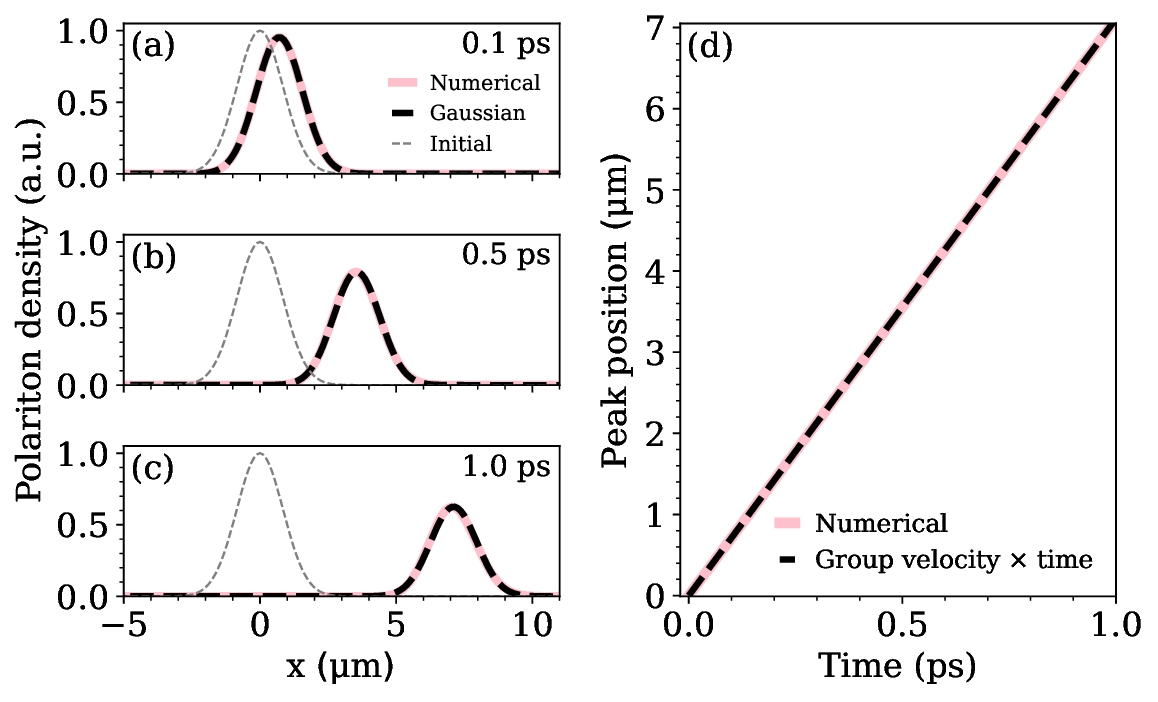}
\caption{Ballistic transport of exciton polaritons in the absence of phonon scattering. (a)-(c) Slices of the angle- and momentum-resolved polariton density at three successive times. Evaluated for a momentum of $Q=1.43 \ \mu\text{m}^{-1}$ (corresponding to the maximum of the group velocity $v_Q^P$), assuming no phonon scattering and a $+10$ meV detuned cavity. The thin dashed grey line shows the initial Gaussian distribution, while the pink curve is the numerical result from solving Eq.~\ref{eq:boltzmann} with $\Gamma_{nQ,mQ}=0$. The dashed black line shows the analytical result (Eq.~\ref{eq:time_dependent_gaussian}). The decay in the Gaussian's height with time is due to radiative decay. (b) The peak position of the Gaussian extracted from the first moment, $\braket{x}_Q$ (pink line), and calculated analytically with $v^P_Q t$ (dashed black line). 
\label{fig:fig_S2}}
\end{figure}

\subsection{Pure ballistic polariton transport}
In the absence of phonon scattering, $\Gamma_{nQ,mQ'}=0$, the Boltzmann equation reduces to a purely ballistic equation that can be solved independently for each momentum (i.e., a collection of two-dimensional advection equations). In Fig.~\ref{fig:fig_S2}(a), the ballistic motion of a single momentum ($1.43 \ \mu\text{m}^{-1}$, corresponding to maximum $v_Q^P$) and angular component of the polariton Wigner function is shown at three different times. It is a shape-preserving evolution; it remains a Gaussian at all times, with only a change in amplitude due to radiative decay and a shift in the offset. Therefore, it is exactly described by a time-dependent Gaussian (dashed black line):
\begin{equation}
N_Q(x,y=0,t) \propto\exp\left(-\frac{(x-v^P_Qt)^2}{2\sigma_0^2}\right)\exp(-\gamma_Q t) \label{eq:time_dependent_gaussian}.
\end{equation}
In Fig.~\ref{fig:fig_S2}(b), the peak position of the Gaussian is calculated with the first moment (pink line), $\braket{x}_Q=\int dx \ x N_Q(x,y=0,t)/\int dx N_Q(x,y=0,t) $,
which can be compared to the analytical result (dashed black line) for the time-dependent offset, $v_Q^P t$, where perfect agreement is found. This illustrates that, without phonon scattering, polaritons move ballistically at the group velocity.

\subsection{Momentum-resolved diffusion coefficient}
\begin{figure}[t!]
\includegraphics[width=\columnwidth*3/5]{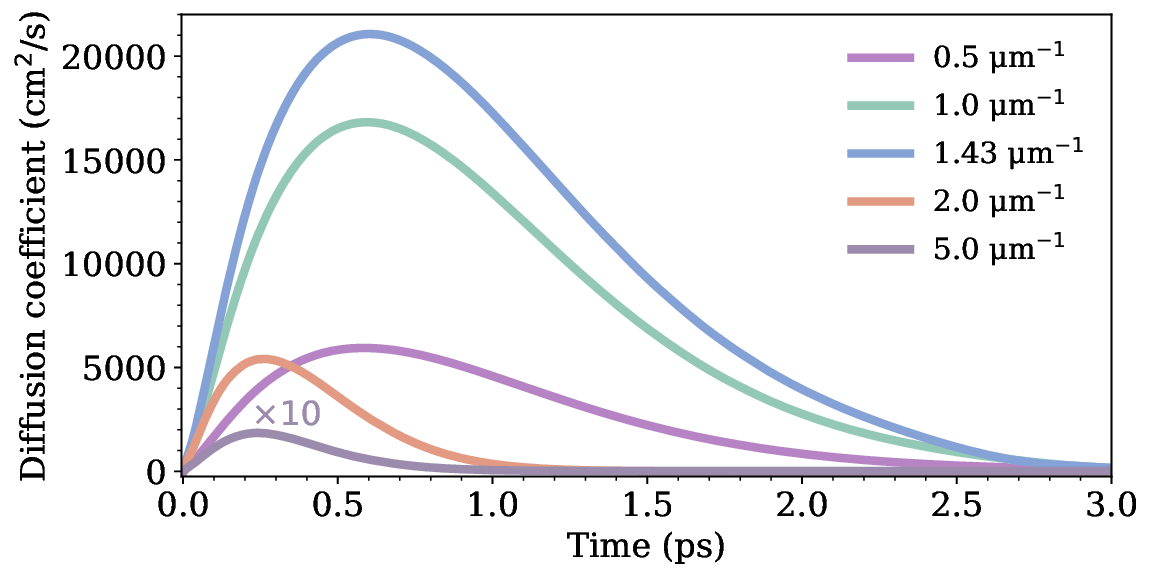}
\caption{ Momentum-resolved effective diffusion coefficients at $300$ K and $\Delta=+10$ meV. Note that the diffusion coefficient for $5$ $\mu$m$^{-1}$ (lavender curve) is multiplied by $10$ for clarity. The transient superdiffusion is found to increase with group velocity, peaking at $1.43$ $\mu$m$^{-1}$, before then dropping significantly for momenta where the acoustic K phonon scattering channel is open, $Q>1.55$ $\mu$m$^{-1}$.
\label{fig:fig_S3}}
\end{figure}

Figure~\ref{fig:fig_S3} shows the momentum-resolved effective diffusion coefficients (i.e., the time derivative of the broadening, $D(t)=\partial_t\sigma^2(t)/4$) at $300$ K and $\Delta=+10$ meV for the five representative momenta considered in Fig.~2(f) of the main text (dashed coloured lines in Figs.~2(b) and (c)). For the first three momenta considered ($0.5$, $1.0$ and $1.43$ $\mu$m$^{-1}$), the peak diffusion increases with the group velocity (see Fig.~2(b)), but the time at which the maximum is reached is roughly the same for all three momenta. This is a consequence of the approximately constant phonon scattering rate over this range of momenta (see the inset of Fig.~2(c)). The effective diffusion coefficients for the next two momenta ($2$ and $5$ $\mu$m$^{-1}$) exhibit a different behaviour. The peak diffusion value is shifted to earlier times and reduced in magnitude due to the opening of the acoustic phonon scattering channel into KK$^\prime$ excitons (see Fig.~2(c)). In particular, the diffusion coefficient at $5$ $\mu$m$^{-1}$ is very small compared to the other momenta due to a combination of increased phonon scattering and reduced group velocity.

\begin{figure}[b!]
\includegraphics[width=\columnwidth*3/5]{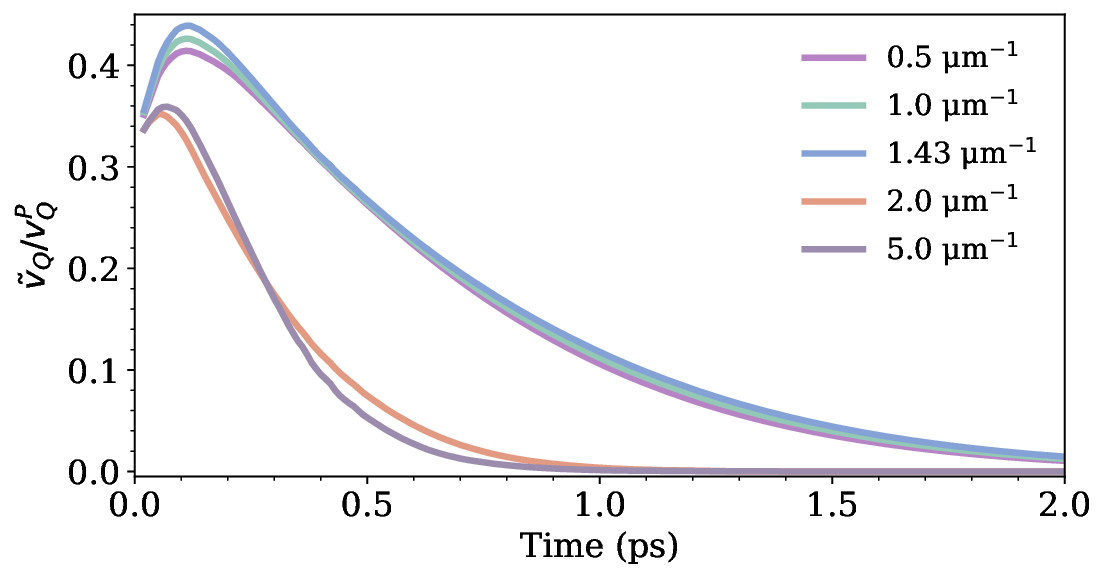}
\caption{Momentum- and angle-resolved effective velocity, divided by $v^P_Q$, at $300$ K and $\Delta=+10$ meV. The effective velocity shows two distinct behaviours depending on the momentum relative to the intervalley scattering opening at $Q=1.55$ $\mu$m$^{-1}$. Components with momenta below this value peak at a slightly higher fraction of the group velocity and decay more slowly over time.
\label{fig:fig_S4}}
\end{figure}

\subsection{Momentum-resolved effective polariton velocity}
Figure~\ref{fig:fig_S4} shows the effective velocity of the momentum- and angle-resolved polariton densities, normalized by $v^P_Q$, for the same five representative momenta as discussed above. Following the same logic, the momentum-resolved results can be divided into two groups based on whether the given momentum falls before or after the opening of the acoustic intervalley scattering pathway between polaritons and KK$^\prime$ excitons. Velocity components with momenta below the opening of the scattering channel ($0.5$, $1.0$ and $1.43$ $\mu$m$^{-1}$) have a slightly increased effective velocity (normalized to $v^P_Q$) and a much slower decay than those above ($2$ and $5$ $\mu$m$^{-1}$).

\subsection{Polariton relaxation}
\begin{figure}[t!]
\includegraphics[width=\columnwidth*3/5]{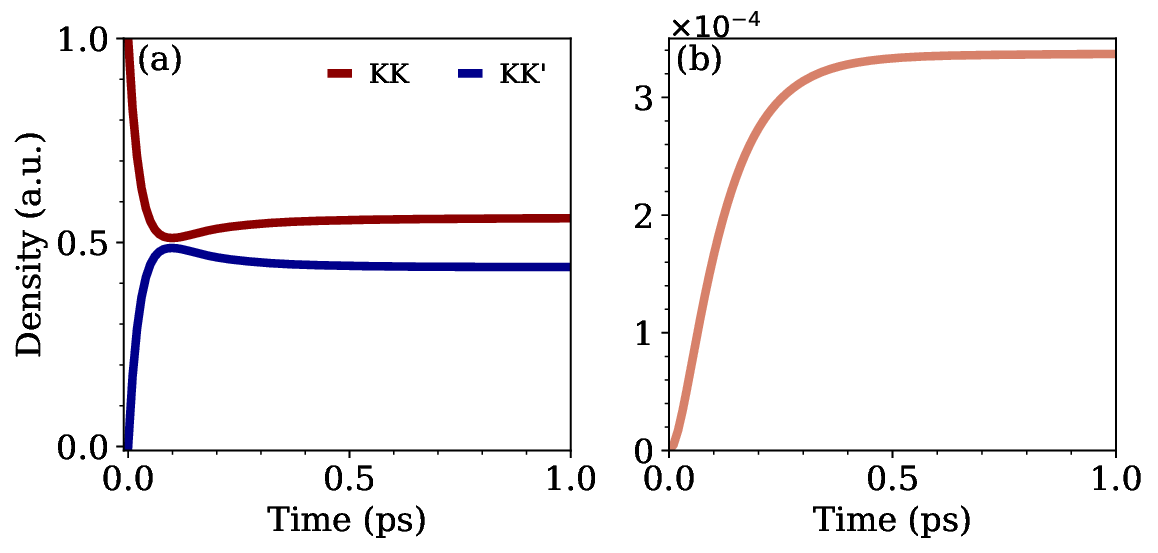}
\caption{Relaxation of the integrated polariton occupation at $r=0$, $T=300$ K and $\Delta=+10$ meV. (a) Time evolution of the KK and KK$^\prime$ density (occupation integrated over all momenta, including both in and outside the lightcone). (b) Time evolution of the polariton density, calculated by integrating the KK occupation over just the lightcone. Note that the densities in this case are on the order of $10^{-4}$ smaller than in (a). 
\label{fig:fig_S5}}
\end{figure}

To better understand the spatiotemporal dynamics shown in the main text, we discuss here the relaxation in momentum space, resolved at a single spatial point $R=0$, at $T=300$ K and $\Delta=+10$ meV. Figure~\ref{fig:fig_S5}(a) shows the time evolution of the integrated occupation over the entire KK and KK$^\prime$ valleys (including polaritons states in the KK lightcone). Over about $100$ ps, there is a rapid transfer of population from the initial hot-exciton distribution in the KK valley to the KK$^\prime$ valley, driven by efficient scattering via phonon emission. Over the next few hundred femtoseconds, the system thermalizes with the occupation in both valleys smoothly evolving into Boltzmann distributions. This is consistent with previous studies of purely excitonic systems \cite{selig2016excitonic,selig2018dark,rosati2021non}. Crucially, we find only a weak bottleneck effect; therefore, a significant polariton population can build up in the lightcone on the same timescale as the exciton reservoir thermalizes. This is illustrated in Fig.~\ref{fig:fig_S5}(b), where the increase in the polariton density (calculated by integrating the KK occupation over only the lightcone) is shown. It is this ultrafast feeding of the polariton population that drives the ballistic-like regime observed in the initial expansion period of the polariton cloud (see Figs.~2-4 in the main text). 

\subsection{Detuning study of cavity-enhanced exciton reservoir broadening}

\begin{figure}[t!]
\includegraphics[width=\columnwidth*3/5]{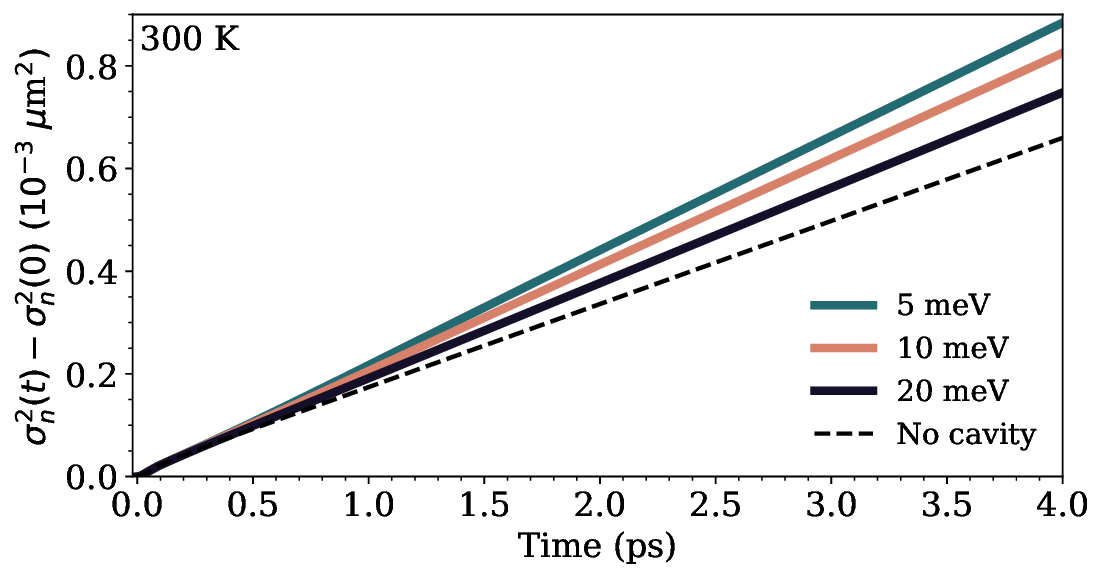}
\caption{Detuning study of the cavity-enhanced total exciton reservoir broadening at $T=300$ K. The black dashed line shows the broadening for a bare monolayer with no cavity. At steady state, a clear detuning dependence of the rate of expansion is observed. Exciton polaritons with a more photonic character lead to a stronger cavity enhancement.
\label{fig:fig_S6}}
\end{figure}

Figure~\ref{fig:fig_S6} presents a detuning study of the total exciton-polariton broadening at $300$ K, i.e., the variance calculated for all states inside and outside the lightcone. Similar to the temperature-dependent study in Fig.~5(a) of the main text, the transient broadening in the first $0.5$ ps is approximately the same for all detuning values and the bare monolayer (dashed black line). This indicates a hot-exciton effect unaffected by the presence of a cavity. At steady state, however, we observe a clear detuning dependence of the rate of expansion. As the cavity is blue detuned ($\Delta$ becomes more positive), polaritons exhibit increased exciton-like character, consequently weakening the cavity enhancement of the expansion.
We confirmed that the enhanced expansion observed in a cavity is characteristic of the entire exciton reservoir, rather than being solely attributable to the rapidly expanding polariton cloud. This was established by an additional study that excluded contributions from exciton polaritons within the lightcone, integrating only states outside it. The results from this closely mirrored those obtained from a full integration.

\subsection{Dependence on initial excitation conditions} \label{sec:init}

\begin{figure}[b!]
\includegraphics[width=\columnwidth*3/5]{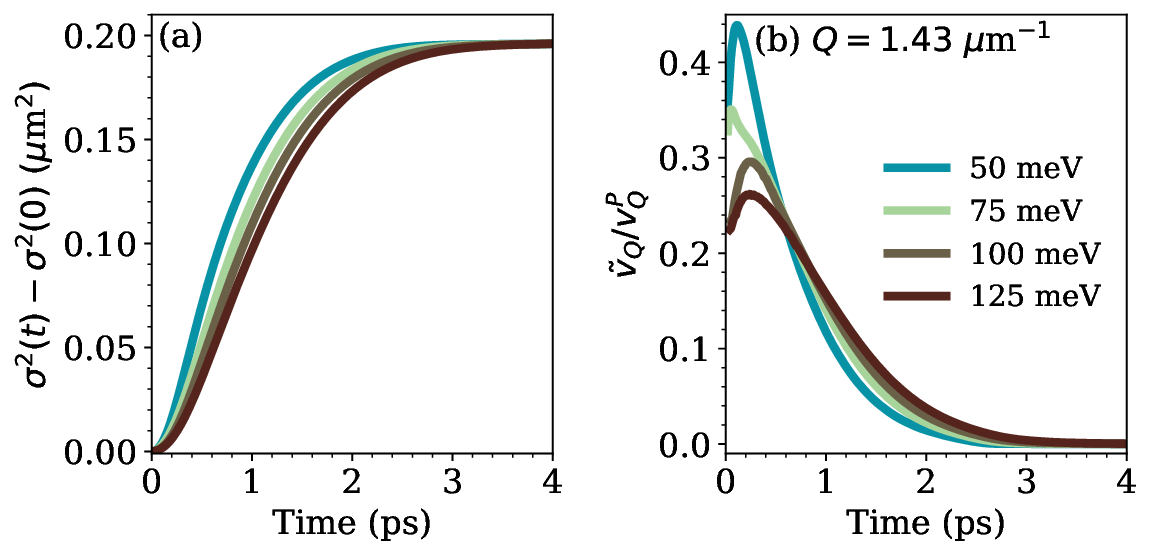}
\caption{Dependence on the initial hot-exciton distribution energy. (a) Polariton broadening shown for different energies of the initial exciton occupation at $300$ K and $\Delta=10$ meV. (b) Corresponding effective velocity of the momentum- and angle-resolved polariton density at $Q=1.43$ $\mu$m$^{-1}$.
\label{fig:fig_S7}}
\end{figure}

Figure~\ref{fig:fig_S7} illustrates the impact of the energy of the initial exciton occupation, that is, the centre energy of the Gaussian distribution of excitons in the KK exciton reservoir at $t=0$. In all preceding results, this was set to $50$ meV with a width of $1$ meV. By changing this, we can modify the ultrafast thermalization period, i.e., the first few hundred femtoseconds where we observe a ballistic-like expansion of the polariton cloud. In reality, it will also  depend on the exact form of non-resonant excitation and the nature of the material's high-energy landscape. Figure~\ref{fig:fig_S7}(a) shows the polariton broadening (integrating over just the lightcone) for a system at $300$ K and $\Delta=10$ meV, examining how it varies for four different initialization energies. We observe that a higher initial KK occupation leads to a slower rate of polariton expansion in real space, indicating a smaller transient effective diffusion. This is attributed to a reduced in-scattering rate from the initial polariton population, which less effectively compensates for the polariton-phonon scattering, resulting in a decreased rate of expansion. This only impacts the transient expansion rate; the spatial width of the polariton cloud in the thermalized limit is found to be approximately the same across all four initialization energies. Interestingly, this dependence on the non-resonant excitation energy is opposite to that observed for bare excitons. Hot excitons with excess kinetic energy diffuse faster than lower-energy excitons \cite{cordovilla2019hot}, enhancing the transient real-space expansion. For the total broadening (integrating over all states inside and outside the lightcone), we find that the hot-exciton effect dominates the transient expansion, similar to bare excitons.

This behaviour is further confirmed in Fig.~\ref{fig:fig_S7}(b), where the peak effective velocity of the momentum- and angle-resolved polariton density (for $Q=1.43$ $\mu$m$^{-1}$) decreases with higher initial occupation energies. However, hotter initial conditions result in a longer temporal tail, causing a larger effective velocity at later times due to the delayed generation of the polariton population. We also find a weak dependence on the initial Gaussian width: a broader initial exciton distribution leads to a lower effective velocity (not shown). Together, these results highlight that the exact relaxation pathway the system takes after non-resonant excitation can significantly impact the type of transient transport behaviour observed in polaritonic systems.

\section{Temperature dependence of bare exciton diffusion in monolayer MoSe2}
\begin{figure}[t!]
\includegraphics[width=\columnwidth*3/5]{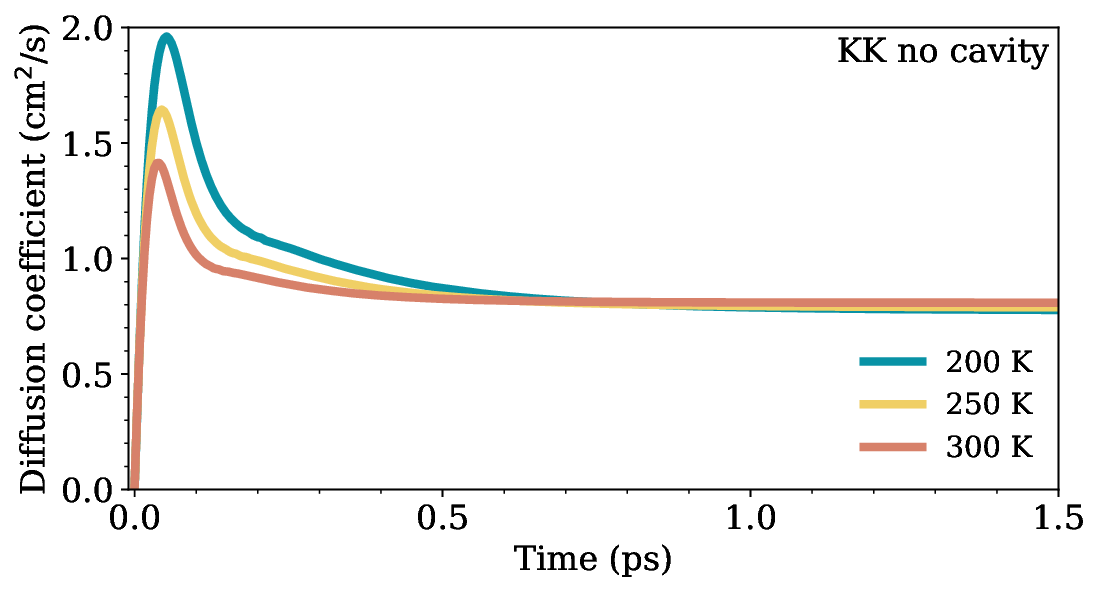}
\caption{Temperature dependence of the effective diffusion coefficient for KK excitons in a bare \ce{MoSe2} monolayer. All three temperatures converge to a steady-state value of approximately $0.8$ cm$^2$/s.
\label{fig:fig_S8}}
\end{figure}

Figure~\ref{fig:fig_S8} illustrates the temperature dependence of the KK effective diffusion coefficient for a bare \ce{MoSe2} monolayer. Beyond the transient regime, characterized by a peak in diffusion around $100$ fs, all three temperatures converge to a steady-state value of approximately $0.8$ cm$^2$/s \cite{rosati2021non,wagner2021nonclassical}. A very similar result is found for the diffusion coefficient of  KK$^\prime$ excitons, which limits towards the same value at thermal equilibrium. At steady state, the temperature insensitivity can be understood from Fick's law as a near-cancellation between competing effects. Specifically, for higher temperatures, the increasing contribution from high-energy exciton occupation via the Boltzmann distribution is offset by the rising exciton-phonon scattering rate.

\section{The polaritonic Fick's law}
In the main text we introduced the modified valley/branch-resolved Fick's law for exciton polaritons,
\begin{equation}
    \tilde{D}_n^{\text{F}}= \frac{\hbar}{2} \sum_\vec{Q}  \frac{\Gamma_{nQ}}{(\gamma_{nQ}+\Gamma_{nQ})^2} v_{nQ}^2 \frac{f_{nQ}}{\sum_{\vec{Q}'}f_{nQ'}}, \label{eq:Ficks}
\end{equation}
where $f_{nQ}/\sum_{\vec{Q}'}f_{nQ'}$ is the Boltzmann distribution normalized with respect to the branch/valley $n$. Equation~\ref{eq:Ficks} provides the steady-state diffusion coefficient for the entire exciton polariton population of that branch/valley, including contributions from states both inside and outside the lightcone. It can be derived by linearizing Eq.~\ref{eq:boltzmann} under the assumption that scattering processes are sufficiently fast that the exciton/polariton occupation only weakly deviates from local quasi-equilibrium \cite{hess1996maxwell}. This means that radiative decay should be weak enough such that it is sensible to talk of a quasi-thermalized system. The additional factor $\Gamma_{nQ}/(\gamma_{nQ}+\Gamma_{nQ})$ in Eq.~\ref{eq:Ficks} accounts for depletion of the steady-state occupation due to radiative decay. This is valid when in-scattering into polaritons is dominated from states outside the lightcone \cite{fitzgerald2024circumventing}.

The \emph{total} diffusion can be found by summing over all branches and valleys yielding
\begin{align}
    \tilde{D}^{\text{F}} &= \frac{\hbar}{2} \sum_n \sum_\vec{Q}  \frac{\Gamma_{nQ}}{(\gamma_{nQ}+\Gamma_{nQ})^2} v_{nQ}^2f_{nQ} \frac{1}{\sum_{m\vec{Q}'}f_{mQ'}} \nonumber
    \\
    &=  \frac{\hbar}{2} \sum_n \underbrace{\sum_\vec{Q}  \frac{\Gamma_{nQ}}{(\gamma_{nQ}+\Gamma_{nQ})^2} v_{nQ}^2 \frac{f_{nQ}}{\sum_{\vec{Q}''}f_{nQ''}}}_{\tilde{D}_n^{\text{F}}} \underbrace{\frac{\sum_{\vec{Q}''}f_{nQ''}}{\sum_{m\vec{Q}'}f_{mQ'}}}_{N_n/N} 
    =
    \sum_{n} \tilde{D}_n^{\text{F}}\frac{N_{n}}{N}
\end{align}
where $N_{n}=\sum_{\vec{Q}}f_{n\vec{Q}}$ and $N=\sum_{n}N_{n}=\sum_{n}\sum_{\vec{Q}}f_{n\vec{Q}}$. Crucially, in the first line we normalized the Boltzmann distribution over all branches and valleys. This reveals that the total steady-state diffusion coefficient is given by the population-weighted sum of valley-resolved diffusion coefficients. 

%\newpage

\bibliography{bib}

\end{document}